\documentclass[twoside,11pt]{article}

\usepackage{amsmath}
\usepackage{graphicx, booktabs, multirow, xcolor, amssymb, epstopdf, enumitem, bm}
\usepackage[ruled]{algorithm2e}

\usepackage{amsmath}
\usepackage{graphicx}
\usepackage{xcolor}
\usepackage{theoremref}
\usepackage{url} 
\usepackage{bbm}
\usepackage[title]{appendix}
\usepackage{subcaption}
\usepackage{caption}
\usepackage{jmlr2e}

\renewcommand{\P}{\mbox{$\mathbb{P}$}}
\newcommand{\E}{\mbox{$\mathbb{E}$}}
\newcommand{\R}{\mbox{$\mathbb{R}$}}

\newcommand{\Var}{\mathrm{Var}}
\newcommand{\Cov}{\mathrm{Cov}}

\newcommand{\1}{\mathbbm{1}}

\let\hat\widehat
\let\tilde\widetilde

\usepackage{bm}
\renewenvironment{proof}{{\bf Proof.}}{$\Box$}

\catcode`@=11
\newskip\beforeproofvskip
\newskip\afterproofvskip
\beforeproofvskip=\medskipamount
\afterproofvskip=\bigskipamount

\def\prooftag{Proof}
\def\proofskip{\enspace}

\def\proof{\@ifnextchar[{\@@proof}{\@proof}}  
\def\@startproof{\par\vskip\beforeproofvskip\leavevmode}
\def\@proof{\@startproof{\scshape\prooftag.}\proofskip}
\def\@@proof[#1]{\@startproof {\scshape\prooftag #1.}\proofskip}

\catcode`@=12

\firstpageno{1}

\begin{document}

\title{Handling Nonmonotone Missing Data with Available Complete-Case Missing Value Assumption}

\author{\name Gang Cheng \email gangc@uw.edu \\
       \addr Department of Statistics\\
       University of Washington\\
       Seattle, WA 98195-4322, USA
       \AND
       \name Yen-Chi Chen \email yenchic@uw.edu \\
       \addr Department of Statistics\\
       University of Washington\\
       Seattle, WA 98195-4322, USA
       \AND
       \name Maureen A.Smith \email maureensmith@wisc.edu \\
       \addr Departments of Population Health Sciences and Family Medicine \\
       University of Wisconsin-Madison \\
       Madison, WI, USA
       \AND
       \name Ying-Qi Zhao \email yingqiz@fredhutch.org \\
       \addr Public Health Sciences Division\\
       Fred Hutchinson Cancer Research Center
       }


\maketitle

\begin{abstract}
Nonmonotone missing data is a common problem in scientific studies.
The conventional ignorability and missing-at-random (MAR) conditions are unlikely to hold for nonmonotone missing data and data analysis can be very challenging with few complete data. 
In this paper, we introduce the available complete-case missing value (ACCMV) assumption 
for handling nonmonotone and missing-not-at-random (MNAR) problems.
Our ACCMV assumption is applicable to 
data set with a small set of complete observations and 
we show that the ACCMV assumption leads to nonparametric identification of the distribution
for the variables of interest.
We further propose an inverse probability weighting estimator, a regression adjustment estimator,
and a multiply-robust estimator for estimating a parameter of interest. 
We studied the underlying asymptotic and efficiency theories of the proposed estimators.
We show the validity of our method with simulation studies and further 
illustrate the applicability of our method by applying it to
a diabetes data set from electronic health records.
\end{abstract}


\section{Introduction}	\label{sec::intro}
\label{sec:intro}

Missing data problems are very common in scientific research \citep{molenberghs2014handbook,little2019statistical}. 
Based on the missing/response patterns, these problems can be categorized into monotone and nonmonotone missing data problems. 
For monotone missing data, variables subject to missing are ordered and if one variable is missing, all subsequent variables are missing. This occurs when individuals drop out of a study, which is common in longitudinal studies \citep{diggle2002analysis}. 
Nonmonotone missingness refers to the case when no such ordering exists \citep{molenberghs2014handbook,little2019statistical}. For example, a participant might drop out and later return to a study. Nonmonotone missingness may also occur for regression analysis when outcomes and predictors are missing under arbitrary patterns. 

Handling nonmonotone missing data is a very challenging task even if we assume missing-at-random (MAR) \citep{robins1997non,sun2018inverse}.
Inverse probability weighting (IPW) estimator for nonmonotone missing data may also be unstable under MAR \citep{sun2018inverse}. Further, \cite{robins1997non} and \cite{vansteelandt2007estimation} have argued that the MAR restriction should not be expected to hold in nonmonotone missing data. 

In this paper, we are interested in dealing with nonmonotone missing data that are missing-not-at-random (MNAR). 
Our study is motivated by an electronic health records (EHRs) data set that contains longitudinal information of diabetes patients.
For patients with diabetes, one important variable is the glycated hemoglobin (HbA1c) measurement and a controlled HbA1c level ($\leq 7\%$) is known to reduce the risk of microvascular complications. However, EHR data also poses significant challenges. EHR data are incomplete as a patient's information is recorded only if and when they visit a clinic. This naturally leads to nonmonotone missing data when a patient reappeared after one or more missed visits. 
Another complication is that the missing patterns of HbA1c are associated with the underlying HbA1c levels. For example, sicker patients with  higher HbA1c levels are likely to visit clinics often and thus have less missing values, while healthier patients are likely to miss visits and thus have more missing values. This suggests that the HbA1c missing mechanism is MNAR. Thus,  we have nonmonotone and MNAR data for the HbA1c measurements. 

The diabetes EHR  data set  contains 8663 patients who were enrolled from 2003 to 2013, and who were followed up every 3 months until the 4th quarter of 2013. Thus, the longest follow up time is 11 years (44 quarters). 
For the purpose of this paper, we will focus on  first-year's data and define $Y_i$ as the HbA1c measurement for the $i$-th quarter with $i = 0, 1, \ldots, 4$ and $Y_0$ as the baseline measurement. There are three main questions we would like to address:
\begin{itemize}
\item {\bf Q1. Single variable of interest.} Given first-year's data ($Y_0,\cdots, Y_4$), 
we are interested in estimating the mean HbA1c levels at the 4-th quarter, 
i.e., $\E[Y_4]$.  

\item {\bf Q2. Multiple variables of interest: summary measures.} Given first-year's data, we want to estimate the probability that a patient successfully controls the HbA1c levels below 7\% for the 3rd and 4th quarters, i.e., 
$P(Y_3 \leq 7\%,Y_4 \leq 7\%)$. Further, we are also interested in estimating the averages of the HbA1c levels for the last two quarters, i.e, $\E[(Y_3 + Y_4)/2]$.

\item {\bf Q3. Multiple variables of interest: marginal parametric model.} Given first-year's data, we want to study the linear relationship between $Y_4$ and $Y_2, Y_3$, i.e., we want to estimate the following linear regression model:
\begin{align*}
  \E[Y_4 | Y_2, Y_3] = \beta_0 + \beta_1 Y_2 + \beta_2 Y_3.
\end{align*}
\end{itemize}

Addressing these questions is a non-trivial problem
because we have
nonmonotone missingness in the data and 
the missingness is MNAR.
Several attemps have been made to handle nonmonotone missing data that is MNAR. One approach is to assume specific parametric models for both the study variables and the missing probability \citep{troxel1998analysis,troxel1998marginal,ibrahim2001missing}. 
Another approach is the no self-censoring or itemwise conditionally independent nonresponse restriction \citep{shpitser2016consistent,sadinle2017itemwise,malinsky2021semiparametric}
and a variant of this idea is the causal graph approach \citep{nabi2020full,mohan2021graphical}. 
\cite{robins1997non} proposed the group permutation model and \cite{zhou2010block} proposed the block conditional MAR model. 
\cite{little1993pattern} and \cite{tchetgen2018discrete} considered the complete-case missing value (CCMV) restriction. \cite{tchetgen2018discrete} used discrete-choice models to generate a class of MNAR assumptions.
\cite{linero2017bayesian} introduced the transformed-observed-data restriction which requires specifying a transformation and it is also a partial identifying restriction. 
\cite{chen2022pattern} introduced the idea of a pattern graph to generate further MNAR assumptions. 
However, all these existing work have limitations and cannot be applied to our problem.
The no self-censoring restriction requires that no variable can be a direct cause of its own missingness status, 
which is unlikely to be true for the diabetes EHR data that we are investigating.
Other methods such as the CCMV and pattern graph rely heavily on the size of the complete cases.
However, for the first year's data $(Y_0, \cdots, Y_4)$, 
complete cases 
only account for $5\%$ of the observations
in our data set.

In this paper, we introduce a useful identifying assumption called available complete-case missing value (ACCMV) assumption for handling nonmonotone missing data that is MNAR. 
In practice 
we often have many variables at hand and only a few of them are of primary interest. We call them \textit{primary variables}.  For those \textit{auxillary variables} that are not of direct interest, they are often correlated with the primary variables and the missing mechanism of the primary variables.
Thus they can be used to assist with the estimation for primary variables.  
For Q1 of the diabetes example, $(Y_0, Y_1, \ldots, Y_3)$ are auxillary variables and $Y_4$ is the primary variable. We can use $(Y_0, Y_1, \ldots, Y_3)$ to help with the estimation of $\E[Y_4]$. 
In such a scenario, the conventional CCMV assumption will require all the variables $(Y_0, \ldots, Y_4)$ to be fully observed for identification. However, requiring auxillary variables to be fully observed is a strong condition to identify parameters that only involve the primary variable.
Ideally, we should also use those observations with primary variables fully observed and auxillary variables partially observed for identification. 

On a high level, the principle of ACCMV imposes an assumption similar to the CCMV on the primary variables for identification and an assumption similar to the available-case missing value (ACMV) assumption \citep{molenberghs1998monotone} on the auxillary variables to improve the effective sample size. 
This allows a much larger set of observations to be used for identification. 
For the diabetes example, close to 48\% of the patients have $Y_4$ observed, while only 5\% of the patients have $Y_0, \ldots, Y_4$ fully observed. Thus, CCMV will only use 5\% of the observation for identification and ACCMV instead will use 48\% of the observations for identification. 
For this reason, ACCMV is particularly suitable for analyzing data sets with few complete cases.

\emph{Outline.}
In Section \ref{sec:notation}, we introduce the relevant notations.
In Section \ref{sec::single}, we study the case with single primary variable. 
We show that ACCMV assumption leads to nonparametric identification of the distributions for the primary variable
and develop an IPW estimator, regression adjustment estimator and a multiply-robust estimator. 
In Section \ref{sec:multiple}, we extend our analysis to multiple primary variables and study the identification, estimation procedure
and efficiency theory. 
We conduct a case study to investigate the scenario of marginal parametric models in Section \ref{sec::MPM}.
Section \ref{sec::sensitivity} studies the problem of sensitivity analysis of the ACCMV assumption.
We conduct simulation studies in Section \ref{sec::simulation} and apply our approach to
the diabetes data set in Section \ref{sec::diabetes}.
All the code for our experiments is available at \url{https://github.com/mathcg/ACCMV}.

\section{Notations}
\label{sec:notation}

In our analysis, we divide all the variables into two sets: a set of variables called \emph{primary variables},  denoted as $L\in\R^d$,
and another set of variables called \emph{auxillary variables}, denoted as $X\in\R^p$.  
We are interested in structures involving the primary variables $L$. 
The auxillary variables $X$ are not of primary interest and mainly help with the estimation involving the primary variables. 
Namely, the parameter of interest is a statistical functional of $L$ and does not involve $X$.
We cannot ignore $X$ in our analysis because $X$ may be related to the missing data mechanism of $L$.
We use $\|\cdot \|$ to denote the $l_2$ norm such that for a vector $x \in \R^d$, we have $\|x\| = \sqrt{ \sum_{i=1}^d x_i^2}$.  
Further, we use $\|\cdot \|_{L_2(P)}$ to denote the $L_2(P)$ norm as $\| f \|_{L_2(P)} =  \left( \int f(x)^2 d P(x) \right)^{1/2}$.


Both $L$ and $X$ are subject to missingness.
We use the binary vector $A\in\{0,1\}^d$ 
to denote the response pattern of $L$, i.e., $A_j=1$ if $L_j$ is observed
and $R\in\{0,1\}^p$  to denote the response pattern of $X$, i.e., $R_j = 1$ if $X_j$ is observed.
We use the notation $X_r = (X_j: r_j = 1)$ and $L_a = (L_j: a_j = 1)$
to denote the observed parts of $X$ and $L$ under pattern $R=r,A=a$. 
Let $1_p = (1,1,\cdots,1)\in\R^p$ and $1_d = (1,1,\cdots,1)\in\R^d$.
We use the notation $\bar r = 1_p-r$ and $\bar a = 1_d-a$
to denote the vector after flipping $0$ and $1$ in $r$ and $a$, respectively. 
The variable $X_{\bar r} = (X_j : r_j=0)$ and $L_{\bar a} = (L_j: a_j = 0)$
will then refer to the missing variables under pattern $R=r$ and $A=a$.
We further define $R \geq r$ if $R_i \geq r_i$ for $i = 1, 2, \ldots, p$. 
For instance, $1010\geq 1000$ but $1010$ cannot be compared with $0100$.

Take the diabetes EHR data as an example.
For question Q1 in Section~\ref{sec::intro},  the primary variable is $L = Y_4\in\R$ and the auxillary variable is $X = (Y_0,\cdots, Y_3)\in\R^4$.
Suppose we only observe $Y_0,Y_2,Y_4$, then this individual would have
response patterns $A =1$ and $R=1010$.
For question Q2, our primary variables is $L = (Y_3,Y_4)\in\R^2$
and the auxillary variables $X= (Y_0,Y_1,Y_2)\in\R^3$.
For the individual who we only observe $Y_0,Y_2,Y_4$, the response pattern is 
$A=01$ and $R=101$.
In what follows, we will give concrete examples
of what ACCMV assumption stands for
in different contexts.


\section{Single Primary Variable for ACCMV: Estimation and Inference} \label{sec::single}
To start with, we consider a simple scenario where 
we only have one primary variable ($d=1$), i.e., $L \in\R$ and $A\in\{0,1\}$,
and we are interested in
estimating the mean functional $\theta = \E(f(L))$
for some known function $f$. 
For the diabetes EHR data, this occurs when we are interested in estimating the average
value of the HbA1c measurement at the end of the first year, i.e., $L = Y_4$ and $\theta = \E(Y_4)$. A straightforward calculation shows that  
\begin{align*}
\theta = \E(f(L)) & = \int f(\ell) p(\ell)d\ell\\
&= \underbrace{\int f(\ell) p(\ell, A=1)d\ell}_{\theta_{1}}
+\sum_{r}\underbrace{\int f(\ell) p(\ell, x_r,R=r, A=0)dx_rd\ell}_{\theta_{0,r}}\\
& = \theta_1 + \sum_r \theta_{0,r}.
\end{align*}
Clearly, $\theta_1$ is identifiable and can be estimated by a simple sample mean, i.e., $\hat \theta_1 = \frac{1}{n}\sum_{i=1}^n f(L_i) I(A_i=1)$,
so we focus on identifying the second term $\theta_{0,r}$.
We can show that 
\begin{align*}
\theta_{0,r}
&= \int f(\ell) p(\ell|x_r,R=r,A=0) p( x_r,R=r,A=0) d\ell dx_r.
\end{align*}
The quantity $p(x_r,R=r,A=0)$
is identifiable from the data.
So the key is to identify the first component $p(\ell | x_r,R=r,A=0)$,
which is also known as the extrapolation density. 

The conventional CCMV assumption
will impose the assumption
\begin{equation}
p(\ell |x_r,R=r,A=0) = p(\ell|x_r,R=1_p,A=1).
\label{eq::ccmv}
\end{equation}
While equation \eqref{eq::ccmv} identifies the parameter $\theta$,
it has a limitation that 
all the information relies on the complete case $R=1_p, A=1$. 
For the diabetes data set, only a very small fraction (5\%)
of the patients have $(Y_0, \ldots, Y_4)$ fully observed.
So the CCMV might lead to an unreliable estimate. 

The ACCMV is based on the insight that
the complete case of $L$
is enough for identifying the parameter of interest and 
we should be more flexible about the response patterns for the auxillary variables $X$.
Formally, the ACCMV assumption imposes the following assumption:
\begin{equation}
p(\ell |x_r,R=r,A=0) = p(\ell|x_r,R\geq r,A=1).
\label{eq::accmv1}
\end{equation}
Namely, to identify $L$  under pattern $R=r, A=0$, 
we use any patterns as long as the primary variable $L$ is observed and the same set of auxillary variables $X_r$ are also observed.
The assumption \eqref{eq::accmv1} allows the use of a much larger
set of observations to infer the information in variable $L$.
We can further prove that ACCMV assumption leads to nonparametric identification \citep{robins2000sensitivity} of the marginal distribution $p(\ell, a)$ and this assumption will not conflict with the observed data. 

\begin{proposition}
Under the ACCMV assumption in equation \eqref{eq::accmv1},
$p(\ell, a)$ is nonparametrically identified. 
\label{prop::NP}
\end{proposition}
It is immediate from Proposition \ref{prop::NP} that $p(\ell)$ is identifiable under the ACCMV assumption. 

\begin{example}
Consider the example where we have $4$ auxillary variables $X_1,X_2,X_3,X_4$
and we focus on the pattern $A=0$ and $R=1010$. 
The CCMV will assume that
$$
p(\ell|x_1,x_3,R=1010, A=0) = p(\ell|x_1,x_3,R=1111, A=1)
$$
and the ACCMV will assume that
$$
p(\ell|x_1,x_3,R=1010, A=0) = p(\ell|x_1,x_3,R\geq 1010, A=1).
$$
For CCMV, the extrapolation density is estimated by observations with $R_i=1111, A_i=1$
whereas in the ACCMV, the extrapolation density is estimated by observations with $R_i \in\{1010, 1110, 1011, 1111\}, A_i=1$. 
Clearly, ACCMV allows us to estimate $p(\ell|x_1,x_3,R = 1010, A=0)$
with a much larger set of observations, leading to a more reliable estimate. 
\end{example}

\subsection{IPW Estimation}	\label{sec::accmv1::ipw}
Instead of directly estimating $p(\ell|x_r,R=r,A=0)$, 
we now propose an IPW
approach to estimate $\theta_{0,r}$.
\begin{lemma} \label{lemma::accmv_ipw}
  The ACCMV assumption \eqref{eq::accmv1} can be equivalently written as follows.
\begin{equation}
\frac{P(R=r,A=0|X_r,L)}{P(R\geq r,A=1|X_r,L)} = \underbrace{\frac{P(R=r,A=0|X_r)}{P(R\geq r,A=1|X_r)}}_{=O_r(X_r)}.
\label{eq::accmv1::ipw}
\end{equation}
\end{lemma}
Lemma \ref{lemma::accmv_ipw} suggests that the ACCMV can be expressed as
requiring the odds 
$P(R=r,A=0|X_r,L)/P(R\geq r,A=1|X_r,L)$
to be independent of the variable $L$. 

An important implication from Lemma \ref{lemma::accmv_ipw}
is that the quantity $O_r(X_r)$ is identifiable 
and we can estimate $O_r(x_r)$ by assuming a parametric model. 
For example, If we set $O_r(x_r; \alpha_r) = \exp(x_r^T \alpha_r)$, the odd can be estimated by
simply
fitting a logistic regression with covariates $X_r$ that
treats pattern $R=r,A=0$ as class 1 and patterns $R\geq r, A=1$ as class 0.
Let $O_r(x_r;\hat \alpha_r)$
be the estimated version of $O_r(x_r)$,
where $\hat \alpha_r$ is the estimated parameter. 

Next, with equation \eqref{eq::accmv1::ipw}, we can rewrite $\theta_{0,r}$ as an identifiable quantity as follows
\begin{equation}
\begin{aligned}
\theta_{0,r}& = \int f(\ell) p(\ell, x_r,R=r, A=0)dx_rd\ell\\
& = \int f(\ell) \frac{p(\ell, x_r,R=r, A=0)}{p(\ell, x_r,R\geq r, A=1)}p(\ell, x_r,R\geq r, A=1) dx_rd\ell\\
& = \int f(\ell) \frac{P(R=r, A=0|\ell, x_r)}{P(R\geq r, A=1|\ell, x_r)}p(\ell, x_r,R\geq r, A=1) dx_rd\ell\\
&\overset{\eqref{eq::accmv1::ipw}}{=}\int f(\ell) O_r (x_r )p(\ell, x_r,R\geq r, A=1) dx_rd\ell\\
& = \E\left(f(L)O_r(X_r) I(R\geq r,A=1)\right).
\end{aligned}
\label{eq::accmv1::ipw1}
\end{equation}
This leads to the following IPW estimator:
\begin{equation*}
\hat \theta_{0,r, \sf IPW} = \frac{1}{n}\sum_{i=1}^n f(L_i)  O_r(X_{i,r}; \hat \alpha_r) I(R_i\geq r, A_i=1).
\end{equation*}
Combining with the estimator $\hat \theta_{1} = \frac{1}{n}\sum_{i=1}^n f(L_i) I(A_i=1)$, 
our final estimator for $\theta$ will be
\begin{equation}
\begin{aligned}
\hat \theta_{ \sf IPW} & = \frac{1}{n} \sum_{i=1}^n f(L_i) I(A_i = 1) \left[1 + \sum_r O_r( X_{i, r}; \hat{\alpha}_r) I(R_i \geq r) \right]
\end{aligned}
\label{eq::accmv1::w2}
\end{equation}
The expression in the last equality shows an elegant form--we can express IPW estimator
as weighting the complete cases $A_i=1$ with weight $1 + \sum_r O_r(X_{i, r}; \hat \alpha_r) I(R_i \geq r)$ and we have the following asymptotic theory for $\hat \theta_{\sf IPW}$. 

\begin{theorem}  \label{thm::thm_ipw_single}
Under the ACCMV assumption in equation \eqref{eq::accmv1::ipw}
and assume that for every $r$,
$$
\sqrt{n}(\hat \alpha_r - \alpha_r^*) = \frac{1}{\sqrt{n}}\sum_{i=1}^n \psi_{r, \alpha_r^*}(X_{i,r}, R_i, A_i) +o_P(1)
$$
for some function $\psi_{r, \alpha_r^*}$ such that $\E[\psi_{r, \alpha_r^*}(X_r, R, A)] = \vec{0}$,  $\E\|\psi_{r, \alpha_r^*} \|^2 < \infty$ and the true odds $O_r(x_r) = O_r(x_r; \alpha_r^*)$.
We assume that $O_r(X_r; \alpha_r)$ is differentiable with respect to $\alpha_r$ and 
\begin{align*}
& \E\|\nabla_{\alpha_r} O_r(X_r; \alpha_r)I(R\geq r) I(A = 1) f(L) \|<\infty \quad 
& \E \| f(L) I(A = 1) O_r(X_r; \alpha_r) I(R \geq r) \|^2 < \infty
\end{align*}
for $\alpha_r\in B(\alpha_r^*, \rho)$ for some $\rho>0$.  
Then 
$$
\sqrt{n}(\hat \theta_{ \sf IPW} -\theta)\overset{d}{\rightarrow} N(0,\sigma^2_{\sf IPW})
$$
for some $\sigma^2_{\sf IPW}>0$.

\end{theorem}
We can compute the variance $\sigma^2_{\sf IPW}$ either through its influence function or use bootstrap.  
More specifically,  we have 
$$
\sqrt{n} (\hat \theta_{\sf IPW} - \theta) = \frac{1}{ \sqrt{n}} \sum_{i=1}^n \phi(X_i, L_i, R_i,  A_i;  \alpha^*) + o_P(1)
$$
and we can estimate $\sigma^2_{\sf IPW}$ with 
\begin{align*}
\hat \sigma^2_{\sf IPW} = \frac{1}{n} \sum_{i=1}^n  (\phi(X_i, L_i, R_i, A_i; \hat \alpha) - \bar{\phi})^2
\end{align*}
where $\bar{\phi} = \frac{1}{n} \sum_{i=1}^n \phi(X_i,  L_i, R_i, A_i;  \hat \alpha)$. 
The form of the influence function $\phi$ can be found in \ref{sec:proof}. 
In practice, we recommend bootstrap for its simplicity. 

Assumptions in \autoref{thm::thm_ipw_single} are mild. 
The asymptotic linear form of $\hat \alpha_r - \alpha_r^*$
is very common when we use a parametric model
and estimate the parameter via the maximum likelihood estimation (MLE).
The condition on the gradient of odds is also very mild.
For conventional methods such as the logistic regression,
this condition holds with covariates that have a bounded second order moment.
The condition on the product of $f(L)$ and odds $O_r(X_r; \alpha_r)$ is also mild. This condition
is required for $\hat \theta_{\sf IPW}$ to have a bounded variance. 
Alternatively we may make the assumption that $O_r(x_r; \alpha_r)$ is bounded by a large constant for any $x_r$ and 
$\alpha_r \in B(\alpha_r^*,  \rho)$. This is very similar to the positivity assumption 
in the IPW literature.


\subsection{Regression Adjustment Estimation}	\label{sec::accmv1::ra}

The ACCMV assumption in equation \eqref{eq::accmv1} leads to the following identification of $\theta_{0,r}$:
\begin{align*}
\theta_{0,r} &=  \int f(\ell) p(\ell|x_r,R=r,A=0) p( x_r,R=r,A=0) d\ell dx_r\\
& \overset{\eqref{eq::accmv1}}{=}  \int f(\ell) p(\ell|x_r,R\geq r,A=1) p( x_r,R=r,A=0) d\ell dx_r\\
&=\int m_{r,0}(x_r) p(x_r,R=r,A=0)dx_r\\
& = \E(m_{r,0}(X_r) I(R=r,A=0)),
\end{align*}
where
\begin{equation}
m_{r,0}(x_r) = \E(f(L)|X_r=x_r,R\geq r, A=1)
\label{eq::accmv1::RA1}
\end{equation}
is the outcome regression model. Thus, we can estimate $\theta_{0,r}$
by imposing a model
$m_{r,0}(x_r) = m_{r,0}(x_r; \beta_r)$
and estimate $\beta_r$ via $\hat \beta_r$
using observations with $R_i\geq r,A_i=1$.
For instance,
we may regress the response $f(L)$ versus covariate $X_r$
from observations with  $R_i\geq r,A_i=1$.
Having estimated $\hat \beta_r$, 
we then construct the estimator
$$
\hat \theta_{0,r, \sf RA} = \frac{1}{n}\sum_{i=1}^n m_{r,0}(X_{i,r}; \hat \beta_r) I(R_i=r, A_i=0)
$$
and the final estimator for $\theta$ will be 
\begin{equation}
\begin{aligned}
\hat \theta_{\sf RA} & = \frac{1}{n}\sum_{i=1}^n [f(L_i) A_i + m_{R_i,0}(X_{i,R_i}; \hat \beta_{R_i}) (1-A_i)].
\end{aligned}
\end{equation}
and we have the following asymptotic theory for $\hat \theta_{\sf RA}$.

\begin{theorem} \label{thm::thm_ra_single}
Under the ACCMV assumption in equation \eqref{eq::accmv1}
and assume that for every $r$,
$$
\sqrt{n}(\hat \beta_r - \beta_r^*) = \frac{1}{\sqrt{n}}\sum_{i=1}^n \psi_{r, \beta_r^*}(L_i, X_{i, r}, R_i, A_i)  +o_P(1)
$$
for some function $\psi_{r, \beta_r^*}$ such that $\E [\psi_{r, \beta_r^*}] = \vec{0}$, $\E \|\psi_{r, \beta_r^*}\|^2 < \infty$
and the true regression function is $m_{r, 0}(x_r) = m_{r, 0}(x_r; \beta_r^*)$.
Also, we assume that $m_{r, 0}$ is differentiable in $\beta_{r}$ and 
\begin{align*}
\E \| \nabla_{\beta_r} m_{r, 0}(X_r; \beta_r) I(R = r, A = 0) \| < \infty \quad \E \| m_{r, 0}(X_r; \beta_r) I(R = r, A = 0) \|^2 < \infty
\end{align*}
for $\beta_r\in B(\beta_r^*, \rho)$ for some $\rho>0$.
Then 
$$
\sqrt{n}(\hat \theta_{ \sf RA} -\theta)\overset{d}{\rightarrow} N(0,\sigma^2_{\sf RA})
$$
for some $\sigma^2_{\sf RA}>0$.


\end{theorem}

Assumptions in  \autoref{thm::thm_ra_single}
takes a similar form as the ones in \autoref{thm::thm_ipw_single}.
These are mild modeling conditions. 
If we use a least square approach to fit the parameter $\beta$
and the true parameter indeed solves the least square equation (occurs when the 
model is correct),
then the asymptotic linear form exists. 
For linear regression, the gradient condition easily holds when the covariates $X$
has bounded second moments.

\subsection{Semi-parametric Theory and Multiply-robust Estimation}

It is known that the IPW and regression adjustment
may not lead to an efficient estimator. 
In this section, we investigate the efficiency theory under the ACCMV assumption.
Since $\theta = \theta_1 + \sum_r \theta_{0,r}$ and the first component is directly identifiable,
we only need to study the efficiency theory of estimating $\theta_{0,r}$.

\begin{theorem} \label{thm::accmv1::eff}
Under the ACCMV assumption, 
the efficient influence function of estimating $\theta_{0,r}$ is
\[
(f(L)-m_{r,0}(X_{r}))O_r(X_r)I(R\geq r, A=1) + m_{r,0}(X_{r}) I(R=r, A=0) -\theta_{0,r}.
\]
\end{theorem}
Based on \autoref{thm::accmv1::eff},
the efficient estimator of $\theta_{0,r}$ is
\begin{equation*}
\hat \theta_{0,r, \sf MR} = \frac{1}{n}\sum_{i=1}^n \left(f(L_i)-\hat m_{r,0}(X_{i,r})\right)\hat O_r(X_{i,r})I(R_i\geq r, A_i=1) +\hat  m_{r,0}(X_{i,r}) I(R_i=r, A_i=0),
\end{equation*}
which leads to the estimator 
\begin{equation}
\hat \theta_{\sf MR} = \sum_r \hat \theta_{0,r, \sf MR}+ \hat \theta_{1}
\label{eq::accmv1::mr}
\end{equation}
where $\hat \theta_{1} = \frac{1}{n}\sum_{i=1}^n f(L_i) I(A_i=1)$.
The two estimated functions $\hat m_{r,0}$ and $\hat O_r$
are the estimators of the regression function $m_{r,0}$ and odds  $O_r$.
We may use the same estimators as in Section~\ref{sec::accmv1::ipw} and $\ref{sec::accmv1::ra}$.
We make the following technical assumptions:

{\bf Assumptions:}
\begin{itemize}
\item[\bf (S1)] For each $r$,  $\hat O_r$ is in a Donsker class ${\cal F}_r$ and $\hat m_{r, 0}$ is in another Donsker class ${\cal G}_r$.  There exist functions $O_r^*(x_r)$ and $m_{r, 0}^*(x_r)$ such that 
\[
\|\hat O_r - O_r^*\|_{L_2(P)} = o_P(1) \quad \| \hat m_{r, 0} - m_{r, 0}^*\|_{L_2(P)} = o_P(1)
\]
\item[\bf (S2)] $f(\ell),  m_{r, 0}(x_r),  O_r(x_r),  \hat m_{r, 0}(x_r), \hat O_r(x_r)$ are uniformly bounded by a large constant $M > 0$ for all $r,  x_r, \ell$. 
\end{itemize}

Assumption (S1) states that estimators $\hat m_{r, 0}$ and $\hat O_r$ should converge to fixed functions.  The Donsker condition is a common condition that controls the complexity of the estimators. 
Assumption (S2) is a technical condition and can be relaxed by stronger moment conditions on each function.  
$O_r(x_r)$ being bounded is related to the positivity assumption in the IPW literature and it is sensible to have $\hat O_r$ 
also bounded as it estimates $O_r(x_r)$.  Further, (S2) holds when these functions are smooth and $X, L$ stay in compact sets. 
The estimator $\hat \theta_{\sf MR}$ 
has the following multiply-robustness properties.

\begin{theorem}
Under the ACCMV assumption, (S1), (S2) 
and appropriate assumptions for ${\cal F}_r$ and ${\cal G}_r$ that we define in \ref{sec::mr_proof}, the estimator $\hat \theta_{\sf MR} $ in equation \eqref{eq::accmv1::mr} satisfies the following properties:
\begin{itemize}

\item {\bf Consistency.} $\hat \theta_{\sf MR} \overset{P}{\rightarrow}\theta$
when 
$$
\sum_{r}\|\hat m_{r, 0} - m_{r, 0}\|_{L_2(P)} \|\hat O_{r} -  O_{r}\|_{L_2(P)}  = o_P(1).
$$

\item {\bf Asymptotic normality.}
$
\sqrt{n}(\hat \theta_{\sf MR} - \theta) \overset{d}{\rightarrow} N(0, \sigma_{\sf eff}^2)
$
when
$$
\sqrt{n} \sum_{r}\|\hat m_{r, 0} - m_{r, 0}\|_{L_2(P)}\|\hat O_{r} -  O_{r}\|_{L_2(P)} = o_P(1).
$$
The quantity $\sigma_{\sf eff}^2$ is the efficiency bound.
\end{itemize}

\label{thm::mr1}
\end{theorem}
The first statement in Theorem~\ref{thm::mr1}
states that as long as for each pattern $r$, 
either the regression estimator $\hat m_{r, 0}$ or the odds estimator $\hat O_r$
is consistent,
the estimator $\hat \theta_{\sf MR}$ will be consistent.
This is known as the multiply-robust property.
The second statement states that if both nuisance functions ($m_r$ and $O_r$) are correctly specified and 
can be estimated sufficiently fast for all patterns $r$,  the final estimator will be asymptotically normal
and achieve the efficiency bound.
The Donsker conditions might be relaxed if sample splitting is employed for estimation of $\hat O_r$ and $\hat m_{r, 0}$. 

Further, if we can assume that both $m_{r, 0}(x_r)$ and $O_r(x_r)$ are parametric functions,  we are able to obtain asymptotic 
normality as long as either $m_{r, 0}^*(x_r) = m_{r, 0}(x_r)$ or $O_{r}^*(x_r) = O_r(x_r)$ for each $r$. 

\begin{corollary}
Under the ACCMV assumption and assuming that $m_{r, 0}(x_r)$ and $O_r^*(x_r)$ are parametric functions for all $r$.  We further
assume that 
\begin{align*}
\| \hat m_{r, 0}(x_r;  \hat \beta_r) - m_{r, 0}(x_r; \beta_r^*) \|_{L_2(P)} = o_P(1) \quad
\| \hat O_r(x_r; \hat \alpha_r) - O_r(x_r; \alpha_r^*) \|_{L_2(P)} = o_P(1) 
\end{align*} 
Then if $m_{r, 0}(x_r; \beta_r^*) = m_{r, 0}(x_r)$ or $O_r(x_r;  \alpha_r^*) = O_r(x_r)$ for each $r$, we have 
$$
\sqrt{n} (\hat \theta_{\sf MR} - \theta) \overset{d}{\rightarrow}  N(0, \sigma^2)
$$
When $m_{r, 0}(x_r; \beta_r^*) = m_{r, 0}(x_r)$ and $O_r(x_r; \alpha_r^*) = O_r(x_r)$ for each $r$, we have $\sigma^2 = \sigma_{\sf eff}^2$. 
\end{corollary}
We can either estimate the variance $\sigma^2$ through the influence functions or use bootstrap.  The form of the influence functions 
can be found in \ref{sec::mr_proof}.  In practice, we recommend using bootstrap to compute the confidence intervals for its simplicity.  

\section{Multiple Primary Variables for ACCMV: Estimation and Inference} \label{sec:multiple}
Now we consider the problem when $L\in\R^d$ is multivariate.
As mentioned before, this occurs when
we are interested in the last two HbA1c measurements for the first year. 
In this case, we have $L= (Y_3,Y_4)$ and $X = (Y_0,Y_1,Y_2)$. We assume that the parameter of interest is $\theta = \E(f(L))$ for some known function $f$. 
Multiple primary variables also occur in the marginal parametric models, which we will
have an in-depth discussion in Section~\ref{sec::MPM}.

When we have multiple primary variables, the complete-case that identifies the variable $L$
will be $A=1_d$. 
Thus, for $a \neq 1_d$,  the ACCMV assumption in equation \eqref{eq::accmv1} will be revised as
\begin{equation}
p(\ell_{\bar a}|\ell_a,x_r,A=a,R=r) = p(\ell_{\bar a}|\ell_a,x_r,A=1_d,R\geq r),
\label{eq::accmv2}
\end{equation}
which is equivalent to
\begin{equation}
\frac{P(R=r,A=a|x_r,\ell)}{P(R\geq r,A=1_d|x_r,\ell)} = \underbrace{\frac{P(R=r,A=a|x_r,\ell_a)}{P(R\geq r,A=1_d|x_r,\ell_a)}}_{= O_{r,a}(x_r,\ell_a)}.
\label{eq::accmv2::ipw}
\end{equation}
Equation \eqref{eq::accmv2::ipw}
is the multivariate version of equation \eqref{eq::accmv1::ipw}.

\begin{proposition}
Under the ACCMV assumption in equation \eqref{eq::accmv2},
$p(\ell, a)$
is nonparametrically identified for any $a \neq 1_d$. 
\label{prop::NP2}
\end{proposition}
Proposition \ref{prop::NP2} shows that the ACCMV assumption for multiple primary variables
nonparametrically identifies the marginal density $p(\ell)$.
So it is an assumption on the missing data without putting any constraints on the observed data. 
Our goal is to identify $\theta$ when our data 
is a collection of IID random elements $(R_i,A_i, X_{i, R_i}, L_{i, A_i})$
for $i=1,2,\cdots, n$.


\subsection{IPW Estimation}	\label{sec::accmv2::ipw}
For any function $f(\ell)$, we have 
\begin{align*}
\theta = \E (f(L)) = \sum_{r, a} \E \left( f(L) I(A = a, R = r) \right) = \sum_{r, a} \theta_{r, a}
\end{align*}
When $a = 1_d$,  $\theta_{r, a} = \E \left( f(L) I(A = a, R = r) \right)$ is identifiable.  When $a \neq 1_d$,  through similar derivations as in \eqref{eq::accmv1::ipw1}, we have 
\begin{align*}
\theta_{r, a} \overset{\eqref{eq::accmv2::ipw}}{=} \E(f(L) O_{r,a}(X_r,L_a) I(A=1_d,R\geq r))
\end{align*}
and the right hand side is clearly identifiable as long as we can estimate $O_{r,a}$. 

Moreover, we have the following equality holds, 
\begin{align*}
\sum_{r,a \neq 1_d} O_{r,a}(X_r,L_a) I(A=1_d,R\geq r) &=\sum_r I(A=1_d, R=r) \sum_{\tau\leq r, a \neq 1_d}O_{\tau, a} (X_\tau,L_a)\\
&= \sum_r Q_{r}(X_r,L) I( R=r, A=1_d),
\end{align*}
with 
\begin{equation}
Q_r(X_r, L) = \sum_{\tau \leq r, a \neq 1_d} O_{\tau, a}(X_{\tau}, L_a).
\label{eq::Q}
\end{equation}
We can then rewrite the above equality as 
\begin{align*}
\theta = \E(f(L))& = \sum_{r,a \neq 1_d}\E(f(L) O_{r,a}(X_r,L_a) I(A=1_d,R\geq r)) + \sum_r \E(f(L) I(A = 1_d, R = r)) \\
& = \E\left(f(L) \sum_r [1 + Q_{r}(X_r,L)] I(R=r, A=1_d)\right).
\end{align*}
The quantity $1 + Q_r(X_r, L)$
behaves like the weight of observation with $A=1_d, R=r$.

Based on the above analysis, our estimation procedure of $\theta$ will be the
following three-step approach: 

\begin{enumerate}
\item {\bf Step 1: estimating individual odds $O_{r,a}$.}
We first estimate $\hat O_{r,a}(X_r,L_a)$ for $a \neq 1_d$. 
This can be done with a simple logistic regression, i.e.,
$\hat O_{r,a}(X_r,L_a) = \exp(\hat \alpha_{r,a}^T(X_r,L_a))$
where $\hat \alpha_{r, a}$ is estimated by comparing pattern $(R=r,A=a)$ versus $(R\geq r, A=1_d)$
using variables $X_r, L_a$.

\item {\bf Step 2: computing total weights $Q_{r}$.}
For each pattern $(R=r,A=1_d)$, we compute
\begin{equation}
\hat Q_{r} (X_r,L) = \sum_{\tau \leq r} \sum_{a \neq 1_d} \hat O_{\tau,a} (X_\tau, L_a).
\label{eq::Q::est}
\end{equation}

\item {\bf Step 3: applying the IPW approach.}
The final estimator is 
\begin{equation}
\hat \theta_{\sf IPW} = \frac{1}{n}\sum_{i=1}^n f(L_i)[ \hat Q_{R_i}(X_{i,R_i},L_i) + 1] I(A_i=1_d).
\label{eq::accmv2::ipw::est}
\end{equation}

\end{enumerate}

\begin{theorem}
Under the assumption \eqref{eq::accmv2::ipw}
and assume that for every $r$ and $a\neq 1_d$,
$$
\sqrt{n}(\hat \alpha_{r,a} - \alpha_{r,a}^*) = \frac{1}{\sqrt{n}}\sum_{i=1}^n \psi_{r,a}(X_{i,r}, L_{i,a}, R_i, A_i) +o_P(1)
$$
for some function $\psi_{r,a}$ such that $\E[\psi_{r, a}] = \vec{0}$ and $\E \|\psi_{r, a}\|^2 < \infty$. The true odds $O_{r,a}(x_r, \ell_a) = O_{r,a}(x_r, \ell_a; \alpha_{r,a}^*)$.
We assume that $O_{r, a}(x_r, \ell_a;  \alpha_{r, a})$ is differentiable with respect to $\alpha_{r, a}$ and 
\begin{align*}
& \E\| \nabla_{\alpha_{r, a}} O_{r, a}(X_r, L_a; \alpha_{r, a}) I(R \geq r, A = 1_d) f(L) \| < \infty  \\
& \E \| f(L) I(R \geq r, A = 1_d) O_{r, a}(X_r, L_a; \alpha_{r, a}^*)\|^2 < \infty
\end{align*}
for $\alpha_{r,a}\in B(\alpha_{r,a}^*, \rho)$ for some $\rho>0$.
Then 
$$
\sqrt{n}(\hat \theta_{ \sf IPW} -\theta)\overset{d}{\rightarrow} N(0,\sigma^2_{\sf IPW})
$$
for some $\sigma^2_{\sf IPW}>0$.
\label{thm::ipw2}
\end{theorem}

The conditions in Theorem~\ref{thm::ipw2} are very similar to the single primary variable case (\autoref{thm::thm_ipw_single}).
The difference is that here we have multiple  response patterns of $L$ that we need to consider.
Again, assuming the logistic regression model (log odds is linear) is correct,
then all these assumptions hold whenever $X$ and $L$ have bounded second moments. 
The proof can be found in \ref{sec::multiple} and the variance can be estimated either through the influence function or bootstrap. 

\subsection{Regression Adjustment Estimation}

Similar to the case of single primary variable scenario,
we may apply a regression adjustment approach to estimate $\theta$ as well. 
The idea is based on the pattern mixture model formulation in equation \eqref{eq::accmv2}
that links the extrapolation density to an observed density. 

Specifically,  for $a \neq 1_d$, 
Equation \eqref{eq::accmv2} implies that 
the parameter $\theta_{r,a} = \E(f(L)I(R=r,A=a))$ can be expressed via the following form:
\begin{align*}
\theta_{r, a} \overset{\eqref{eq::accmv2}}{=}\E(m_{r,a}(X_r,L_a) I(R=r,A=a))
\end{align*}
where 
\begin{equation}
m_{r,a}(X_r,L_a) = \E(f(L)|L_a, X_r, R\geq r, A=1_d),
\label{eq::accmv2::RA1}
\end{equation}
is the outcome regression model.
As a result, the regression adjustment approach leads to the following two-stage estimator of $\theta$:
\begin{enumerate}
\item {\bf Step 1: estimating the outcome regression.}
For each $r,a$ with $a\neq 1_d$,
we estimate $m_{r,a}(X_r,L_a)$ via an estimator $\hat m_{r,a}(X_r,L_a)$
using observations with $R\geq r, A=1_d$ and variables $L, X_r$. 
This can be done by placing a parametric model $m_{r,a}(X_r,L_a; \beta_{r,a})$
and estimating the underlying parameter $\hat \beta_{r,a}$. 

\item {\bf Step 2: regression adjustment.} 
With the estimates from step 1, our final estimate will be 
\begin{equation}
\hat \theta_{\sf RA} = \frac{1}{n}\sum_{i=1}^n  \left[f(L_i)I(A_i=1_d) + \hat m_{R_i,A_i}(X_{i,R_i}, L_{A_i}) I(A_i\neq 1_d)\right].
\label{eq::accmv2::RA2}
\end{equation}

\end{enumerate}

The regression adjustment estimator can be interpreted as follows.
When we have a complete observation of the primary variable ($A_i=1_d$),
we observe $f(L_i)$. 
When any entries of $L$ is missing, we find a proper model $\hat m_{R, A}$
based on the response pattern in $L$, together
with the response pattern in $X$, 
and compute the predicted value of $f(L)$.

\begin{theorem}
Under the assumption of equation \eqref{eq::accmv2}
and assume that for every $r,a\neq 1_d$,
$$
\sqrt{n}(\hat \beta_{r,a} - \beta_{r,a}^*) = \frac{1}{\sqrt{n}}\sum_{i=1}^n \psi_{r,a}(X_{i,r}, L_{i,a}, R_i, A_i) +o_P(1)
$$
for some function $\psi_{r,a}$ such that $\E[\psi_{r, a}] = 0$ and $\E \|\psi_{r, a}\|^2 < \infty$. Further, assume that the true regression $m_{r,a}(x_r, \ell_a) = m_{r,a}(x_r, \ell_a; \beta_{r,a}^*)$,   $m_{r, a}$ is differentiable in $\beta_{r, a}$ and 
\begin{align*}
& \E\|\nabla_{\beta_{r,a}} m_{r,a}(X_r, L_a; \beta_{r,a})I(R =  r, A=a)\|<\infty \\
& \E \| m_{r, a}(X_r, L_a; \beta_{r, a}) I(R = r, A = a) \|^2 < \infty
\end{align*}
for $\beta_{r,a}\in B(\beta_{r,a}^*, \rho)$ for some $\rho>0$.
Then 
$$
\sqrt{n}(\hat \theta_{ \sf RA} -\theta)\overset{d}{\rightarrow} N(0,\sigma^2_{\sf RA})
$$
for some $\sigma^2_{\sf RA}>0$.

\label{thm::ra2}
\end{theorem}

Conditions in Theorem~\ref{thm::ra2} is similar to the conditions in \autoref{thm::thm_ra_single}
except that $L$ is multivariate. 
The modeling conditions are also mild; linear regression models will satisfy them
when we have bounded second moments of both $X$ and $L$. 
The proof can be found in \ref{sec::multiple} and the variance can be computed either 
based on the influence functions or bootstrap.

\subsection{Semi-parametric Theory and Multiply-robust Estimation}
Both IPW and regression adjustment are known to be inefficient.
To improve the efficiency of the estimator,
we first derive the efficient influence function of $\theta_{r, a}$.

\begin{theorem}\label{thm::accmv2::eff}
Under the ACCMV assumption in equation \eqref{eq::accmv2::ipw},
the efficient influence function of estimating $\theta_{r, a}$ when $a\neq 1_d$ is
$$
[f(L)-m_{r,a}(X_{r}, L_a)]O_{r,a}(X_r, L_a)I(R\geq r, A=1_d) + m_{r,a}(X_{r}, L_a) I(R=r, A=a) -\theta_{r, a}.
$$
\end{theorem}
The above theorem implies that we can construct an efficient estimator using the following approach:
\begin{equation}
\begin{aligned}
\hat \theta_{\sf MR} &= \frac{1}{n}\sum_{i=1}^n \left[ \sum_{r,a \neq 1_d} \{ [ f(L_i)-\hat m_{r,a}(X_{i,r})] \hat O_{r,a}(X_{i,r}, L_{i,a})I(R_i\geq r, A_i=1_d) \right.\\
&\left. \qquad+ \hat m_{r,a}(X_{i,r}, L_{i,a}) I(R_i=r, A_i=a) \} + f(L_i) I(A_i = 1_d)\right],
\end{aligned}
\label{eq::accmv2::mr}
\end{equation}
where $\hat O_{r,a}$ and $\hat m_{r,a}$ are 
estimators of the odds $O_{r,a}$ in equation \eqref{eq::accmv2::ipw} and the outcome regression $m_{r,a}$ in equation \eqref{eq::accmv2::RA1},
respectively.  We make the following technical assumptions: 
\textbf{Assumptions: }
\begin{itemize}
\item[\bf (M1)] For each $r$ and $a \neq 1_d$, $\hat O_{r, a}$ is in a Donsker class ${\cal F}_{r, a}$ and $\hat m_{r, a}$ is in a Donsker class ${\cal G}_{r, a}$.
There exist functions $O_{r, a}^*(x_r, l_a)$ and $m_{r, a}^*(x_r, l_a)$ such that 
\begin{align*}
\| \hat O_{r, a} - O_{r, a}^*\|_{ L_2(P)} = o_P(1)  \quad \| \hat m_{r, a} - m_{r, a}^* \|_{ L_2(P)} = o_P(1)
\end{align*}
\item[\bf (M2)] $f(\ell), m_{r, a}(x_r, l_a), O_{r, a}(x_r, l_a),  \hat m_{r, a}(x_r, l_a), \hat O_{r, a}(x_r, l_a)$ are uniformly bounded by a large constant $M > 0$ for all $x_r, \ell, r$ and $a \neq 1_d$. 
\end{itemize}
Assumptions (M1) and (M2) are multivariate versions of (S1) and (S2). 
\begin{theorem}
Under the ACCMV assumption \eqref{eq::accmv2},  (M1), (M2) and appropriate assumptions for ${\cal F}_{r, a}$ and ${\cal G}_{r, a}$ that we define in \ref{sec::multiple}, 
the estimator $\hat \theta_{\sf MR}$
has the following properties:
\begin{itemize}

\item {\bf Consistency.} $\hat \theta_{\sf MR} \overset{P}{\rightarrow}\theta$
when 
$$
\sum_{r,a\neq 1_d}\|\hat m_{r,a} - m_{r,a}\|_{L_2(P)}\|\hat O_{r,a} -  O_{r,a}\|_{L_2(P)}  = o_P(1).
$$

\item {\bf Asymptotic normality.}
$
\sqrt{n}(\hat \theta_{\sf MR} - \theta) \overset{d}{\rightarrow} N(0, \sigma_{\sf eff}^2)
$
when
$$
\sqrt{n} \sum_{r,a\neq 1_d}\|\hat m_{r,a} - m_{r,a}\|_{L_2(P)}\|\hat O_{r,a} -  O_{r,a}\|_{L_2(P)} = o_P(1).
$$
The quantity $\sigma_{\sf eff}^2$ is the efficiency bound.
\end{itemize}
\label{thm::accmv::mr}
\end{theorem}

Theorem~\ref{thm::accmv::mr} implies that the estimator $\hat \theta_{\sf MR}$
is multiply-robust
in the sense that as long as we have either $m_{r,a}$ or $O_{r,a}$ being consistently estimated
for all $r$ and $a\neq 1_d$, 
the estimator $\hat \theta_{\sf MR}$ will be consistent.
Further it achieves the efficiency bound when the two sets of nuisance models are estimated sufficiently fast.

\section{Multiple Primary Variables for ACCMV: Marginal Parametric Model}	\label{sec::MPM}

In practice, we often impose a marginal parametric model over the primary variable $L$
and use the data to estimate the underlying parameter. 
To start with, we consider two motivating examples. 

\begin{example}
{\bf (Modeling the marginal distribution)}
We assume that $L\sim p(\ell;\theta^*)$, where $p(\cdot; \theta)$ is a known parametric distribution such as a multivariate Gaussian,
and the goal is to estimate the underlying parameter $\theta^*$. 
A typical approach to estimate $\theta^*$ is the maximum likelihood estimator (MLE).
Under usual regularity conditions,
the true parameter solves the population score equation:
$$
\theta^*: 0 = \E(s(\theta^*|L)), \qquad s(\theta|\ell) = \nabla_\theta \log p(\ell;\theta).
$$
When there is no missingness in $L$, the MLE is obtained from the following sample score equation:
$$
\hat \theta_{\sf MLE}: 0 = \frac{1}{n}\sum_{i=1}^n s(\hat\theta_{\sf MLE}|L_i).
$$
To give a concrete example of this,
consider again the one-year diabetes data $Y_0,\cdots, Y_4$.
Suppose that we are interested in the joint distribution of the last two visits, i.e., $L = (Y_3,Y_4)$, 
and we assume that it follows a bivariate Gaussian, i.e., $p(\ell;\theta) = p(y_3,y_4; \mu, \Sigma)$,
where $\mu\in\R^2$ is the mean vector and $\Sigma\in\R^{2\times 2}$ is the covariance matrix.
Then we can easily estimate $\mu$ and $\Sigma$ using the MLE. 
\end{example}

\begin{example}
{\bf (Modeling the marginal moment restricted model)}
It is also very common that the parameter of interest
may be a moment restricted model among variables in $L$.
For instance, we may impose a linear model $\E(L_1|L_{-1}) = L^T_{-1}\theta^* $,
where $L_{-1}$ is all variables in $L$ except the first one (for simplicity, we ignore the intercept).
The parameter of interest is the regression coefficient $\theta^*$. 
In this case, we often estimate the parameter $\theta^*$
by the least square approach, i.e.,  at the population level, the parameter $\theta^*$ satisfies
$$
\theta^* = {\sf argmin}_\theta \E((L_1- L^T_{-1}\theta^*)^2),
$$
or equivalently, the parameter $\theta^*$ solves the following equation:
$$
\vec{0} = \E(L_{-1}(L_1 - L^T_{-1}\theta^*)).
$$
When there is no missingness in $L$, the least square estimate solves the following estimating equation 
$$
\hat \theta_{\sf LS}: \vec{0} = \frac{1}{n}\sum_{i=1}^n L_{i,-1} (L_{i,1} - L^T_{i,-1}\hat \theta_{\sf LS}).
$$
In the diabetes data, if we are interested in the linear relationship among $Y_2,Y_3$ and $Y_4$, we can use the model above and treat $Y_4 = L_1$ and $(Y_2,Y_3) = L_{-1}$. 
\end{example}

In both examples, we see that the parameter of interest is now defined through
a population estimating equation 
\begin{equation}
0 = \E(s(\theta^*|L)).
\label{eq::accmv2::EE}
\end{equation}
So we will focus on the case of parameters defined through an estimating equation, and how to obtain a consistent estimate
when there are missingness in $L$ based on the ACCMV assumption 
\eqref{eq::accmv2::ipw}.

%
%
%


\subsection{IPW Marginal Parametric Model}


The IPW approach in Section~\ref{sec::accmv2::ipw}
can be easily adopted to the marginal parametric model. 
Specifically, the population estimating equation of \eqref{eq::accmv2::EE}
can be written as
\begin{equation}
\begin{aligned}
0 & = \E(s(\theta^*|L))\\
& \overset{\eqref{eq::accmv2::ipw}}{=} \sum_{r, a \neq 1_d}\E(s(\theta^*|L) O_{r,a}(X_r,L_a) I(A=1_d,R\geq r)) + \sum_r \E(s(\theta^* | L) I(A = 1_d, R = r)\\
& = \E\left(s(\theta^*|L) \left[ \sum_{r, a \neq 1_d}O_{r,a}(X_r,L_a) I(A=1_d,R\geq r)  + \sum_r  I(A = 1_d, R = r)\right]
\right) \\
& = \E\left(s(\theta^*|L) I(A=1_d)\sum_{r}[Q_{r}(X_r,L) + 1] I(R= r)\right),
\end{aligned}
\label{eq::accmv2::ipw3}
\end{equation}
where the weight function $Q_r$ is from equation \eqref{eq::Q}.

As a result, the three-step procedure in Section~\ref{sec::accmv2::ipw} can be applied
here with a mild modification:
\begin{enumerate}
\item {\bf Step 1: estimating individual odds $O_{r,a}$.}
For $r$ and $a \neq 1_d$, 
we first estimate $\hat O_{r,a}(X_r,L_a)$. 
This can be done by  a simple logistic regression, i.e.,
$\hat O_{r,a}(X_r,L_a) = O_{r,a}(X_r,L_a;\hat \alpha_{r, a})$
where $\hat \alpha_{r, a}$ is estimated by comparing pattern $(R=r,A=a)$ versus $(R\geq r, A=1_d)$
using variables $X_r, L_a$.

\item {\bf Step 2: computing total weights $Q_{r,a}$.}
For each pattern $(R=r,A=1_d)$, we compute its total weight
$$
\hat Q_{r} (X_r,L) = \sum_{\tau \leq r} \sum_{a \neq 1_d} \hat O_{\tau,a} (X_\tau, L_a).
$$
\item {\bf Step 3: solving the weighted estimating equation.}
The final estimator $\hat\theta$ is from
\begin{equation}
\hat \theta: 0 = \sum_{i=1}^n s(\theta | L_i) [\hat Q_{R_i}(X_{i,R_i},L_i) + 1] I(A_i=1_d).
\label{eq::accmv2::ipw4}
\end{equation}
\end{enumerate}
The first two steps are the same as Section~\ref{sec::accmv2::ipw}.
We only need to modify the last step
by solving a weighted estimating equation. 
We have the following asymptotic results for $\hat \theta$. 
\begin{theorem}
Under assumption \eqref{eq::accmv2::ipw}
and assume that for every $r$ and $a\neq 1_d$,
$$
\sqrt{n}(\hat \alpha_{r,a} - \alpha_{r,a}^*) = \frac{1}{\sqrt{n}}\sum_{i=1}^n \psi_{r,a}(X_{i,r}, L_{i,a}, R_i, A_i) +o_P(1)
$$
for some function $\psi_{r,a}$ such that $\E[\psi_{r, a}] = 0$ and $\E \|\psi_{r, a}\|^2 < \infty$. The true odds $O_{r,a}(x_r, \ell_a) = O_{r,a}(x_r, \ell_a; \alpha_{r,a}^*)$.  Next we assume that $O_{r, a}$ is differentiable in $\alpha_{r, a}$ and 
$$
\E \|
s(\theta^* | L) \nabla_{\alpha_{r, a}} O_{r, a} (X_r, L_a; \alpha_{r, a}) I(A = 1_d, R \geq r) \| < \infty
$$
for $\alpha_{r,a}\in B(\alpha_{r,a}^*, \rho)$ for some $\rho>0$.
Further we assume that 
$$
I(\theta^*) = \E\left[ \nabla_{\theta} s(\theta | L) |_{\theta = \theta_0} \left[ \sum_{r, a \neq 1_d} O_{r, a}(X_r,  L_a; \alpha_{r, a}^*) I(A = 1_d, R \geq r) + I(A = 1_d) \right]  \right]
$$
exists and is invertible. Assuming that $\hat \theta \rightarrow_p \theta^*$, we have  
$$
\sqrt{n}(\hat \theta  -\theta^*)\overset{d}{\rightarrow} N(0,  \Sigma)
$$
for some covariance matrix $\Sigma$.
\label{thm::ipw_margin_parametric_model}
\end{theorem}

\subsection{Potential Problems with Regression Adjustment}

The marginal parametric model
has one distinct property from
the general case of multiple primary variables:
the regression adjustment 
method and the multiply-robust approach
may be problematic. 
The main reason is: both regression-adjustment and multiply-robust approach
will involve imposing a conditional model on one subset of $L$ conditioned on another subset of $L$. 
This procedure implicitly places a model constraint
on the distribution of $L$,
which will conflict with the 
marginal parametric model
when the model is not designed well. 
The multiply-robust estimator also suffers from the same problem
since it involves a model on the outcome regression. 
On the other hand, the  odds in the IPW approach is a conditional model
on the  
selection odds $P(R=r,A=a|x_r,\ell_a) / P(R\geq r,A=1_d|x_r,\ell_a)$,
so it is always compatible with the marginal parametric model.
Hence, we recommend using the IPW approach 
in the case of  marginal parametric model.

This phenomenon is similar to the model congeniality problem
introduced in \cite{meng1994multiple}. 
The model congeniality problem refers to the case where the imputation model
may not be compatible with the analysis model imposed on the imputed data. 
An imputation model can be viewed as a Monte Carlo approximation to the regression adjustment
method and the marginal parametric model on $L$ is the analysis model in \cite{meng1994multiple}. 
Thus, the model conflicting problem we encounter when using regression adjustment on marginal parametric model
can be viewed as another form of model congeniality problem.

\section{Sensitivity Analysis via Exponential Tilting} \label{sec::sensitivity}

The fact that the ACCMV assumption is nonparametrically identified implies that it cannot be tested by the data,
which means that it is a weak assumption.
However, it is possible that the ACCMV assumption 
may not be correct or is only approximately correct.
The sensitivity analysis \citep{little2012prevention} is often conducted
to study how the estimate changes when we slightly perturb the underlying missing data assumption. 

Here  we propose to perform sensitivity analysis of ACCMV via an exponential tilting approach \citep{kim2011semiparametric,shao2016semiparametric,zhao2017semiparametric}.
For $a \neq 1_d$, recall that the ACCMV in equation \eqref{eq::accmv2::ipw} requires:
$$
\frac{P(R=r,A=a|x_r,\ell)}{P(R\geq r,A=1_d|x_r,\ell)} = \underbrace{\frac{P(R=r,A=a|x_r,\ell_a)}{P(R\geq r,A=1_d|x_r,\ell_a)}}_{= O_{r,a}(x_r,\ell_a)}.
$$

In reality,  the odds on the left-hand-side of the above equality
may depend on the unobserved value of $L$. 
Using the concept of exponential tilting, we propose to perturb assumption \eqref{eq::accmv2::ipw} as follows:
\begin{equation} 
\frac{P(R=r,A=a|x_r,\ell)}{P(R\geq r,A=1_d|x_r,\ell)} = O_{r,a}(x_r,\ell_a) \cdot \exp(\delta^T_{\bar a}\ell_{\bar a}),
\label{eq::accmv2::SA1}
\end{equation}
where $\delta_{\bar a}\in\R^{|\ell_{\bar a}|}$ is a given vector
that represents the amount of perturbation from the ACCMV assumption.
Clearly, when $\delta_{\bar a}$ is a zero vector,
equation \eqref{eq::accmv2::SA1} reduces to the usual ACCMV assumption.

In practice, we will choose a sensitivity parameter vector $\delta\in\R^d$ first,
which implies $\delta_{\bar a}$ for every $a$.
Then based on the perturbation \eqref{eq::accmv2::SA1},
we compute the modified final estimate.
For the IPW estimator, 
We only need to change 
\[
\hat Q_r(X_r,L) = \sum_{\tau\leq r} \sum_{a \neq 1_d} \hat O_{r,a}(X_\tau, L_a)
\]
in equation \eqref{eq::Q::est}
to
$$
\tilde Q_r(X_r,L;\delta) = \sum_{\tau\leq r} \sum_{a \neq 1_d} \hat O_{r,a}(X_\tau, L_a) \exp(\delta_{\bar a}^T L_{\bar a})
$$
and change the final estimator in equation \eqref{eq::accmv2::ipw::est}
to
\begin{equation}
\tilde \theta_{\sf IPW, \delta} = \frac{1}{\tilde n}\sum_{i=1}^n f(L_i) [\tilde Q_{R_i}(X_{i,R_i},L_i;\delta) + 1] I(A_i=1_d).
\label{eq::accmv2::ipw::est::SA}
\end{equation}
with $\tilde n = \sum_{i=1}^n [\tilde Q_{R, i}(X_{i, R_i}, L_i; \delta) + 1]I(A_i = 1_d)$. 
Note that $\hat O_{r, a}(X_\tau, L_a)$ is estimated under assumption \eqref{eq::accmv2::ipw}.

Under a logistic regression model, the sensitivity parameter in the exponential tilting approach \eqref{eq::accmv2::SA1}
has a nice interpretation. 
Recall that the logistic regression model will model 
$$
O_{r,a}(x_r,\ell_a) = \exp(\alpha_{r,a}^T(x_r,\ell_a)).
$$
Thus, equation \eqref{eq::accmv2::SA1} will become
$$
O_{r, a}(x_r, \ell) = \frac{P(R=r,A=a|x_r,\ell)}{P(R\geq r,A=1_d|x_r,\ell)} = \exp\left(\alpha_{r,a}^T(x_r,\ell_a) + \delta^T_{\bar a}\ell_{\bar a}\right).
$$
Each $\delta_j$ and each element $\alpha_{r,a,j}$ have the same interpretation--they are
the coefficient on linear model of the log odds. 
Consider a specific example that $L= (L_1,L_2)\in\R^2$, $a=10$ ($L_1$ is observed) and the coefficient $\alpha_{r,a}$ on $L_1$ is $2$.
Then a sensitivity parameter $\delta_{01}=1$ can be interpreted as \emph{the effect of the unobserved variable $L_2$ on the log odds 
is half of the estimated effect of the observed variable $L_1$.}
Thus, practitioners can use this as a way to think about a feasible range of the sensitivity parameter $\delta$.

\section{Simulation Study}	\label{sec::simulation}
We now show the validity of our methods with simulation studies. Section \ref{simulation::single} considers the case when there is a single primary variable. Section \ref{simulation::multiple} considers the case when there are multiple primary variables. Section \ref{simulation::mpm} applies our ACCMV assumption to a linear regression model. 

\subsection{Single Primary Variable} \label{simulation::single}
We have $L = Y_3$ and $X = (Y_1, Y_2)$ and we are interested in estimating $\theta = \E[Y_3]$. 
Let $|r| = \sum_r r_i$ be the number of observed variables. 
Next, we generate data as follows:
\begin{enumerate}
\item $(L, X_r) | A = 1, R = r \sim N(\mu_{|r| + 1}, \Sigma_{|r| + 1})$
\item $X_r | A = 0, R = r \sim N(\mu_{|r|}, \Sigma_{|r|})$
\end{enumerate}
with $\mu_1 = 1$, $\mu_2 = (1, -1)^T$, $\mu_3 = (0, -1, -1)^T$ and 
  $$
  \Sigma_1 = 1 \ \text{,} \ \Sigma_2 = \left(\begin{array}{cc} 1 & 1/2 \\ 1/2 & 1 \end{array}\right) \ \text{ and } \Sigma_3 = \left(\begin{array}{ccc} 1 & 1/2 & 1/2 \\
  1/2 & 1 & 1/2 \\
  1/2 & 1/2 & 1 \end{array}\right)
  $$
Further, we assume that $P(A = j, R = r) = 1/8$ for $j = 0, 1$ and $r \in \{00, 01, 10, 11\}$. Note that under ACCMV assumption,  $L | X_r, R = r, A = 0$ for $r \in \{00, 01, 10, 11\}$ are also specified given the data generations above. 

We first consider estimation using the regression adjustment method. 
Under ACCMV, we can compute that $\theta = \E[Y_3] = \frac{89}{96}$ and the details are left in \ref{sec::app::simulation}.  
We fit linear regression models
$$
m_{r, 0}(x_r; \beta_r) = \E[Y_3 | X_r = x_r, R \geq r, A = 1; \beta_r]
$$
for $r \in \{00, 01, 10, 11\}$ and get $\hat \beta_r$.  
The form of the linear regression models can be found in \ref{sec::app::simulation}. 
Then, we can get the estimates using regression adjustment as 
\begin{align*}
  \hat \theta_{\sf RA} = \frac{1}{n} \sum_{i=1}^n \left[Y_{i, 3} I(A_i = 1) + \sum_r m_{r, 0}(X_{i, r}; \hat \beta_r)I(R_i = r, A_i = 0)\right]
\end{align*}

Next, we consider the IPW estimates. 
We can compute the odds functions as follows.
\begin{align*}
  & O_{00} = \frac{P(A = 0, R = 00)}{P(A = 1, R \geq 00)} = \frac{1}{4} \\
  & O_{10}(y_1) = \frac{P(A = 0, R = 10 | y_1)}{P(A = 1, R \geq 10 | y_1)}  = \frac{1}{2}\exp(2y_1) \\
  & O_{01}(y_2) = \frac{P(A = 0, R = 01 | y_2)}{P(A = 1, R \geq 01 | y_2)} = \frac{1}{2} \exp(2 y_2) \\
  & O_{11}(y_1, y_2) = \frac{P(A = 0, R = 11 | y_1, y_2)}{P(A = 1, R = 11 | y_1, y_2)} = \exp\left( 
    \frac{8}{3}y_1 - \frac{4}{3}y_2 - \frac{4}{3}
  \right)
\end{align*}
To get the estimate, we  fit a logistic regression model
\begin{align*}
P(R = r, A = 0 | X_r,  \{R \geq r, A = 1\} \cup \{R = r, A = 0\}; \alpha_r) = \frac{O_r(x_r; \alpha_r)}{1 + O_r(x_r; \alpha_r)}
\end{align*}
for each $r \in \{00, 01, 10, 11\}$ and get $\hat \alpha_r$. Then we can get the estimate using IPW as 
\begin{align*}
\hat \theta_{\sf IPW} = \frac{1}{n} \sum_i Y_{3, i} I(A_i = 1) \left[ 1 + \sum_r O_r(X_{i, r}; \hat \alpha_r) I(R_i \geq r)
\right]
\end{align*}
For the multiply-robust estimator, we can use the linear regression models and 
logistic regression models that we fitted before. 
Then for each $r$,  we can get the estimate as 
\begin{align*}
\hat \theta_{0,r, \sf MR} &= \frac{1}{n}\sum_{i=1}^n \left(f(L_i) - m_{r,0}(X_{i,r}; \hat \beta_r)\right) O_r(X_{i,r};  \hat \alpha_r)I(R_i\geq r, A_i=1) \\
& + m_{r,0}(X_{i,r}; \hat \beta_r) I(R_i=r, A_i=0),
\end{align*}
Our final multiply-robust estimator is then 
\begin{equation}
\hat \theta_{\sf MR} = \sum_r (\hat \theta_{0,r, \sf MR}+ \hat \theta_{1,r})
\end{equation}
where $\hat \theta_{1,r} = \frac{1}{n}\sum_{i=1}^n f(L_i) I(A_i=1, R_i=r)$.  

For the multiply-robust estimator, we first consider the case when all the regression functions and odds functions are correctly specified. 
Next we consider the case when one of the regression function is misspecified. When $r = 11$, the correct regression model is $m_{11, 0}(x_r; \beta_{11}) = \beta_0 + \beta_1 Y_1 + \beta_2 Y_2$ and 
we fit a linear regression model with $Y_1$ only.  We also apply the same model misspecification to the regression adjustment estimator.  
We further consider the case when one of the odds function is mis-specified. When $r = 11$, the correct odds function is $O_{11}(x_r; \alpha_{11}) = \alpha_0 + \alpha_1 Y_1 + \alpha_2 Y_2$ and we fit a logistic regression with intercept only. Again we apply the same model misspecification to the IPW estimator.  
Finally we consider the case when both the regression function and the odds function is incorrect. When $r = 11$, we fit a linear regression model with $Y_1$ only and we fit a logistic regression model with intercept only for the odds function. 

Table \ref{table::single_primary_variables} contains the simulation results. We generate 1,000 samples with $n = 2000$. The bias is computed as the difference of the average of 1,000 parameter estimates and the true value of $\E[Y_3]$. Sample standard error (SE) is computed as the standard error of the 1,000 parameter estimates and the mean theoretical SE is computed as the mean of the 1,000 SE estimates. For all three methods, we estimate the SE for the estimators through their corresponding influence functions\footnote{The actual form of the influen functions can be found in Appendix.}. Note that bootstrap is an alternative approach to estimate the SE. The sample standard SE reflects the true SE of the estimator and the mean theoretical SE reflects the accuracy of the SE estimated through the influence functions. CI stands for Confidence Interval, RA stands for regression adjustment and MR stands for multiply-robust.

Based on the simulation results, we can see that 95\% CI of IPW is undercovering when all the odds function are correctly specified. This is primarily due to the under-estimation of the SE for the IPW estimators in this setup. For this specific data generation setting, the difficulty is that we are estimating the variance of a very heavy-tailed distribution. 
We can see that the mean theoretical SE is much smaller than the sample SE, which suggests that the estimated SE is much smaller than the true SE. Next, IPW with misspecified odds function leads to much larger bias and even worse coverage for the 95 CI\%. We also observe that the mean theoretical SE is much smaller than the sample SE. However, multiply-robust estimator with the same misspecified odds function obtains much better performance. 
We can see that the bias is very small, the coverage of the 95\% CI is very close to the nominal coverage and the difference between mean theoretical SE and the sample SE is also much smaller now. 
This shows the robustness of the multiply-robust estimator. 

Further, regression adjustment achieves nominal coverages and smallest SE estimates when all regression functions are correctly specified. RA with misspecified regression function also leads to large bias and bad coverage for the 95\% CI. Again, multiply-robust estimator with the same model misspecification is able to reduce the bias and improve the coverages. We can see that in this case, all multiply-robust estimators underestimate their variances for the same reason as the IPW estimator.
However, the coverages of the 95\% CI are much better for multiply-robust estimators compared to the IPW estimator. When both the regression and odds function are misspecified, the multiply-robust estimator also obtained biased estimates and bad coverages for the 95\% CI. Finally, we also include the results estimating $\E[Y_3]$ using complete-case analysis. It is clear from the simulations that under ACCMV assumption, the data is missing-not-at-random as complete-case analysis is severely biased. 
\begin{table}[h]
  \caption{Simulations results for estimating $\E[Y_3]$ when $n = 2000$}
  \begin{center}
  \begin{tabular}{ c c c c c }
  \hline
  \hline
   Methods   &  Bias  & Sample SE & Mean theoretical SE & Coverage of 95\% CI\\
   \hline
  IPW & -0.006 & 0.216 & 0.113 & 0.778 \\
  IPW (incorrect) & -0.084 & 0.174 & 0.090 & 0.536 \\
  RA & -0.001 & 0.044 & 0.043 & 0.955 \\
  RA (incorrect) & 0.040 & 0.045 & 0.046 & 0.857 \\
  MR (correct) & -0.002 & 0.105 & 0.072 & 0.931 \\
  MR (IPW incorrect) & 0.000 & 0.069 & 0.057 & 0.939\\
  MR (RA incorrect) & -0.000 & 0.118 & 0.074 & 0.920\\
  MR (Both incorrect) & 0.041 & 0.069 & 0.058 & 0.870 \\
  Complete Case & -0.178 & 0.035 & 0.034 & 0.001 \\
  \hline
  \hline
  \end{tabular}
  \end{center}
  \label{table::single_primary_variables}
\end{table}

\subsection{Multiple Primary Variables} \label{simulation::multiple}
We have 
$L = (Y_3, Y_4)$ and $X = (Y_1, Y_2)$.  The parameter of interest is $\theta = \E[Y_3 Y_4]$ and this allows the estimation of the $\Cov(Y_3, Y_4)$ given $\E[Y_3]$ and $\E[Y_4]$. For any $a$, let $|a| = \sum_i a_i$ be the number of observed primary variables.
We generate the data as follows. 
\begin{enumerate}
\item $(L, X_r) | A = 11, R = r \sim N(1_{2 + |r|}, \Sigma_{2 + |r|})$ for $r \in \{00, 01, 10, 11\}$. 
\item $(L_a, X_r) | A = a, R = r \sim N(\mu_{1 + |r|}, \Sigma_{1 + |r|})$ for any $a \in \{01, 10\}$ and any $r \in \{00, 01, 10, 11\}$. 
\item $X_r | A = 00, R = r \sim N(\mu_{|r|}, \Sigma_{|r|})$ for any $r \in \{01, 10, 11\}$. 
\end{enumerate}
where $1_d = (1, \ldots, 1)^T \in \R^d$, $\Sigma_d = 1/2 I_d + 1/2 1_d 1_d^T$ and 
$$
\mu_1 = 0.5 \quad \mu_2 = 1_2 \quad \mu_3 = 1_3.
$$
Further, we also assume that $P(A = a, R = r) = 1/16$ for $a \in \{00, 01, 10, 11\}$ and $r \in \{00, 01, 10, 11\}$. 

We first consider regression adjustment method.  Under the ACCMV assumption, we can compute $\theta = \E[Y_3 Y_4] = 175/128$ and the details can be found in \ref{sec::app::simulation}. 
Now to get the estimate, we need to fit the following regression models:
\begin{align*}
 m_{r, a}(x_r; \beta_{r, a}) = \E[Y_3 Y_4 | X_r = x_r, R \geq r, A = 11; \beta_{r, a}]
\end{align*}
for $a \neq 11$. Note that equivalently we can fit the following regression models
\begin{align*}
& m'_{r, 00}(x_r;  \beta_{r, 00}) = \E[Y_3 Y_4 | X_r = x_r, R \geq r, A = 11; \beta_{r, 00}] \\
& m'_{r, 01}(x_r,  y_4; \beta_{r, 01}) = \E[Y_3 | X_r = x_r, Y_4 = y_4,  R \geq r,  A = 11;  \beta_{r, 01}] \\
& m'_{r, 10}(x_r,  y_3; \beta_{r, 10}) = \E[Y_4 | X_r = x_r, Y_3 = y_3,  R \geq r, A = 11;  \beta_{r, 10}]
\end{align*}
for $r \in \{00, 01, 10, 11\}$ and get $\hat \beta_{r, a}$. The actual form of the regression models can be found in \ref{sec::app::simulation} and we have  
\begin{align*}
  m_{r, a}(x_r, l_a; \beta_{r, a}) = \begin{cases} m'_{r, a}(x_r, l_a; \beta_{r, a}) & a = 00 \\
    m'_{r, a}(x_r, l_a; \beta_{r, a}) l_a & a = 01, 10
  \end{cases}
\end{align*}
Then we can get the estimate using regression adjustment as 
\begin{align*}
\hat \theta_{\sf RA} & = \frac{1}{n} \sum_{i=1}^n [
Y_{3, i} Y_{4, i} I(A_i = 11) + \sum_{r, a \neq 11} m_{r, a}(X_{r, i}, L_{a, i}; \hat \beta_{r, a}) I(R_i = r,  A_i = a)]
\end{align*}
Next, we consider the IPW estimation. It is not hard to get that for $a \neq 11$,
$\log O_{r, a}(X_r, L_a)$ is a linear function of $X_r$ and $L_a$. 
Then we can fit a logistic regression model as follows:
\begin{align*}
  P(R = r, A = a | x_r, l_a, \{R \geq r, A = 11\} \cup \{R = r, A = a\}; \alpha_{r, a}) = \frac{O_{r, a}(x_r, l_a; \alpha_{r, a})}{1 + O_{r, a}(X_r, L_a; \alpha_{r, a})}
\end{align*}
for each $r$,  $a \neq 11$ and get $\hat \alpha_{r, a}$. We can get the estimate using IPW as 
\begin{align*}
  \hat \theta_{\sf IPW} = \frac{1}{n} \sum_{i=1}^n Y_{3, i} Y_{4, i} I(A_i = 11) \left[\sum_{r, a \neq 11} O_{r, a}(X_{r, i}, L_{a, i}; \hat \alpha_{r, a}) I(R_i \geq r) + 1\right]
\end{align*}
We now consider the multiply-robust estimation.  
We can get the estimate using the following multiply-robust estimator
\begin{align*}
  \hat \theta_{\sf MR} & = \frac{1}{n} \sum_{i=1}^n \left[ \sum_{r, a \neq 11} \{
    (f(L_i) - m_{r, a}(X_{i, r}, L_{i, a}; \hat \beta_{r, a})) O_{r, a}(X_{i, r}, L_{i, a}; \hat \alpha_{r, a}) I(R_i \geq r, A_i = 11) \right. \\ 
    & \left. + m_{r, a}(X_{i, r}, L_{i, a}; \hat \beta_{r, a}) I(R_i = r, A_i = a)
    \} + f(L_i) I(A_i = 11)
  \right]
\end{align*}
with the same estimators $m_{r, a}(X_r, L_a; \hat \beta_{r, a})$ and $O_{r, a}(X_r, L_a; \hat \alpha_{r, a})$ as before. 
Again, we consider the case when all the regression functions and odds functions are correctly specified. Next, we consider the case when two of the regression estimators is misspecified. When $R = 00$ and $A = 01$, the true regression model is $\E[Y_3 | Y_4] = \beta_0 + \beta_1 Y_4$ and we fit a linear regression with intercept only. When $R = 00$ and $A = 10$, the true regression model is $\E[Y_4 | Y_3] = \beta_0 + \beta_1 Y_3$ and we also fit a linear regression model with intercept only. We apply the same model misspecification to the regression adjustment estimator. 
Further, we consider the case when two of the odds function are misspecified.  When $R = 00$ and $A = 01$, the true odds function is $O_{r, a}(Y_4) = \exp(\alpha_0 + \alpha_1 Y_4)$ and we fit a logistic regression with intercept only.  When $R = 00$ and $A = 10$, the true odds function is $O_{r, a}(Y_3) = \exp(\alpha_0 + \alpha_1 Y_3)$ and we also fit a logistic regression with intercept only. We also apply the same model misspecification to the IPW estimator. Finally, we consider the case when both the regression function and the odds function are misspecified. When $R = 00$ and $A = 10$, we fit a linear regression model with intercept only for the regression function and we also fit a logistic regression model with intercept only for the odds function. When $R = 00$ and $A = 01$, again we fit linear regression and logistic regression models with intercepts only. 
\begin{table}[h]
  \caption{Simulations results for estimating $\E[Y_3 Y_4]$ when $n = 2000$}
  \begin{center}
  \begin{tabular}{ c c c c c }
  \hline
  \hline
   Methods   &  Bias  & Sample SE & Mean theoretical SE & Coverage of 95\% CI\\
   \hline
  IPW & -0.000 & 0.075 & 0.074 & 0.943 \\
  IPW (incorrect) & 0.078 & 0.079 & 0.079 & 0.852 \\
  RA & -0.001 & 0.065 & 0.066 & 0.956 \\
  RA (incorrect) & -0.048 & 0.065 & 0.068 & 0.892 \\
  MR (correct) & -0.001 & 0.066 & 0.066 & 0.948 \\
  MR (IPW incorrect) & -0.001 & 0.065 & 0.066 & 0.949\\
  MR (RA incorrect) & -0.001 & 0.067 & 0.067 & 0.952\\
  MR (both incorrect) & 0.014 & 0.068 & 0.065 & 0.943 \\ 
  Complete Case & 0.131 & 0.092 & 0.092 & 0.723 \\
  \hline
  \hline
  \end{tabular}
  \end{center}
  \label{table::multiple_primary_variables}
\end{table}

From table \ref{table::multiple_primary_variables}, we can see that IPW, RA and multiply-robust estimators all obtained close to 0 bias and achieves nominal coverages for the 95\% CIs when the regression or odds functions are correctly specified. In comparison, we also observe that IPW estimator obtained relatively large SE estimates, while multiply robust estimators and regression adjustment estimator obtained relatively similar SE estimates. As expected, both IPW and regression adjustment estimators fails when the regression or odds functions are misspecified. Multiply-robust estimators with the same model misspecification are able to achieve nominal coverages and small biases. Further, when both the regression and odds functions are misspecified, multiply-robust estimators obtain relatively large bias. Finally, the complete-case analysis again obtained severely biased estimates. 




\subsection{Marginal Parametric Model} \label{simulation::mpm}
In this section, we consider the following setup for the marginal parametric model.  We have 
$L = (Y_2, Y_3)$ and $X = Y_1$.  We want to estimate the following linear regression model 
\begin{align*}
  \E[Y_3 | Y_2] = \beta_0 + \beta_1 Y_2
\end{align*}
Further, we assume that 
\begin{enumerate}
  \item $(X, L) \sim_d N(\mu_0, \Sigma)$  with $\mu_0 = (1, 0, -1)^T$ and 
$$
\Sigma = \left(\begin{array}{ccc} 1 & 1/2 & 1/2 \\
  1/2 & 1 & 1/2 \\
  1/2 & 1/2 & 1 \end{array}\right)
$$
  \item $P(R = 0, A = a | X, L) = h(X, L) \exp\left(L^T 1_{2} \right)$ for $a \in \{00, 01, 10, 11\}$. 
  \item $P(R = 1, A = a | X, L) = h(X, L) \exp(L^T 1_2 + 0.5X)$ for $a \in \{00, 01, 10\}$.
  \item $P(R = 1, A = 11 | X, L) = h(X, L) \exp(L^T 1_2)$. 
\end{enumerate}
with $h(X, L) = 1 / \left[5 \exp(L^T 1_2) + 3 \exp(L^T 1_2 + 0.5X)\right]$ being the normalization term. 
It can be verified that this data generation satisfies the ACCMV assumption. Given the data generation above, we have 
\begin{align*}
  \E[Y_3 | Y_2] = -1 + \frac{1}{2} Y_2
\end{align*}
As discussed in section \ref{sec::MPM}, we will be using IPW to estimate the parameters for the linear regression model. 
We can estimate individual odds $O_{r, a}$ by fitting a logistic regression model as follows:
\begin{align*}
  P(R = r, A = a | x_r, l_a; \{R \geq r, A = 11\} \cup \{R = r, A = a\}; \alpha_{r, a}) = \frac{O_{r, a}(x_r, l_a; \alpha_{r, a})}{1 + O_{r, a}(x_r, l_a; \alpha_{r, a})}
\end{align*}
and get $\hat \alpha_{r, a}$. 
Next, we compute the total weights for each $r$ such that 
$$
\hat Q_r(X_r, L) = \sum_{\tau \leq r} \sum_{a \neq 11} O_{\tau, a}(X_\tau, L_a; \hat \alpha_{\tau, a})
$$
Finally, we can get $\hat \beta$ by solving the following weighted estimating equation:
\begin{align*}
  \sum_{i=1}^n s(\hat \beta | L_i) \left[\sum_{r} \hat Q_r(X_{i, r}, L_i)I(R_i = r) + 1\right] I(A_i = 11) = 0
\end{align*}
From table \ref{table::marginal_parametric_model}, we can see that IPW gives close to 0 bias and achieves nominal coverage for the 95\% confidence intervals. On the other hand, complete-case analysis obtained biased estimates.

\begin{table}[h]
  \caption{Simulations results for the marginal parametric model when $n = 2000$.}
  \begin{center}
  \begin{tabular}{ c c c c c }
  \hline
   & \multicolumn{2}{c}{Bias (SE)} & \multicolumn{2}{c}{Coverage of 95\% CI} \\
  \hline
   Methods   &  $\beta_0$ & $\beta_1$ & $\beta_0$ & $\beta_1$ \\
   \hline
  IPW & -0.001 (0.039) & -0.001 (0.046) & 0.949 & 0.938 \\
  Complete Case & -0.061 (0.042) & -0.008 (0.043) & 0.725 & 0.943 \\
  \hline
  \hline
  \end{tabular}
  \end{center}
  \label{table::marginal_parametric_model}
\end{table}

\section{Applications to the Diabetes Data}	\label{sec::diabetes}
We apply the proposed estimation procedures to the diabetes EHR data set, assuming that ACCMV holds. 
This data set contains 8663 patients who were followed up every 3 months from 2003 to 2013. 

\subsection{Summary Measures of the HbA1c Levels}
We focus on the HbA1c levels measured from the baseline and the first year, $(Y_0, Y_1, \ldots, Y_4)$. 
We now answer the first two questions raised in the introduction. We estimate 
\begin{enumerate}
\item The mean HbA1c levels at the end of the first year, $\E[Y_4]$.
\item The proportion of patients that have their HbA1c levels controlled, meaning that their HbA1c levels are below 7\%, i.e., $P(Y_3 \leq 7, Y_4 \leq 7)$. \footnote{For convenience, we multiply the value of HbA1c levels by 100.}
\item The averages of the HbA1c levels for the last two quarters, $\E[Y_3 + Y_4]/2$.
\end{enumerate}
For the estimation of $\E[Y_4]$, the primary variable is $L = Y_4$ and the auxillary variables are $X = (Y_0, \ldots, Y_3)$. For the estimations of $\E[Y_3 + Y_4] / 2$ and $P(Y_3 \leq 7, Y_4 \leq 7)$, the primary variables are $L = (Y_3, Y_4)$ and the auxillary variables are $X = (Y_0, Y_1, Y_2)$. 

We construct the 95\% confidence intervals using bootstrap.  The results are given in Table \ref{table::real_data_results} and \ref{table::real_data_results_regression}. We can see that IPW, regression adjustment and multiply robust estimators all obtain quite similar results. The only exception is with the estimation of $P(Y_3 \leq 7, Y_4 \leq 7)$, where IPW and multiply-robust estimators obtain non-overlap 95\% confidence intervals. This could be due to the incorrect specifications of the odds functions.  
We can also see that results of complete-case analysis are different from the rest three methods. 
Complete-case analysis tends to over-estimate the HbA1c levels. 
This suggets that the missingness of HbA1c is relevant to the underlying values of HbA1c, which agrees with the intuition that healthier patients are more likely to miss their HbA1c measurements. 

More specifically, our ACCMV approach shows that both $\E[Y_4]$ and $\E[Y_3 + Y_4] / 2$ is less than 7 and more than 50\% of the participants have their HbA1c levels controlled for the last two quarters; while complete-case analysis obtain completely opposite results. 
This highlights the importance of treating missing data in a real data set. 

\begin{table}[h]
  \caption{Summary statistics computed on diabetes data set}
  \begin{center}
  \begin{tabular}{ c c c c  }
  \hline
  \hline
   Methods   &  $\E[Y_4]$  & $P(Y_3 \leq 7, Y_4 \leq 7)$ & $ \E[Y_3 + Y_4] / 2$ \\
   \hline
  IPW & 6.927 (6.897 - 6.956) & 0.534 (0.515 - 0.556) & 6.931 (6.900 - 6.958)\\
  Regression Adjustment & 6.936 (6.907 - 6.966) & 0.569 (0.554 - 0.583) & 6.955 (6.925 - 6.983)\\
  Multiply Robust & 6.931 (6.902 - 6.960) & 0.575 (0.559 - 0.589) & 6.949 (6.920 - 6.976)\\
  Complete Case & 7.011 (6.974 - 7.049) & 0.454 (0.431 - 0.477) & 7.197 (7.143 - 7.252)\\
  \hline
  \hline
  \end{tabular}
  \end{center}
  \label{table::real_data_results}
\end{table}

\subsection{Sensitivity Analysis}


\begin{figure}
  \centering
  \includegraphics[width=\textwidth]{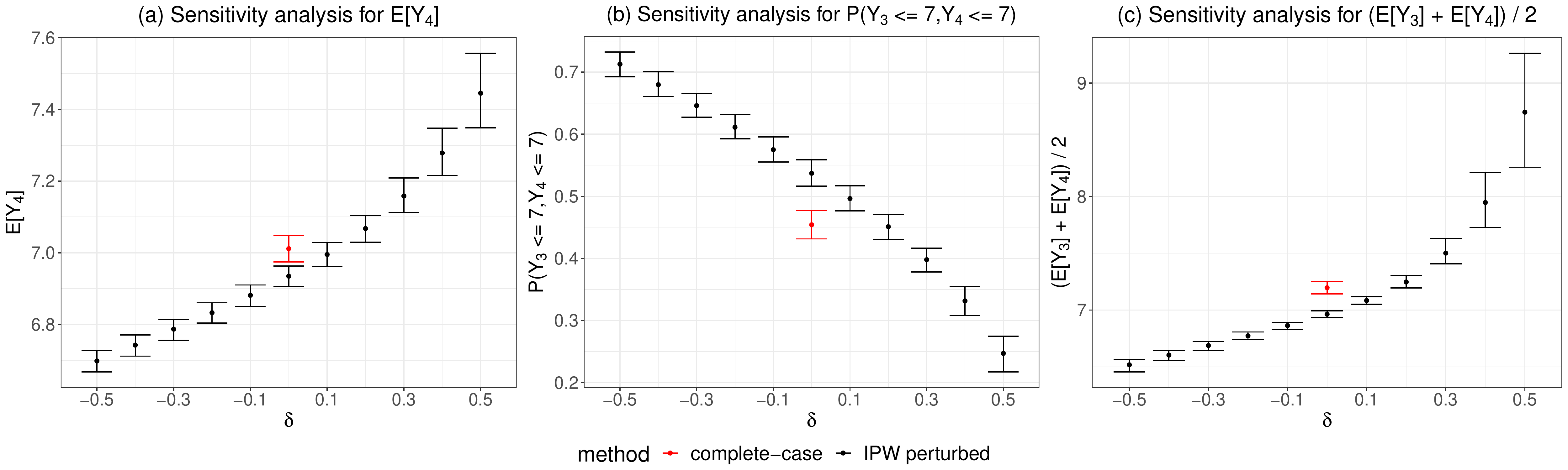}
  \caption{Sensitivity analysis of the ACCMV assumption by exponential tilting. We examine how $\E[Y_4], P(Y_3 \leq 7, Y_4 \leq 7)$ and $\E[Y_3 + Y_4] / 2$ and confidence intervals change with respect to different values of the sensitivity parameter $\delta$.}
  \label{fig::sa_summary_measures}
\end{figure}

We also perform sensitivity analysis by the exponential tilting approach proposed in section \ref{sec::sensitivity}. We use the same sensitivity parameter for all patterns, i.e., every element of $\delta_{\bar a}$ in equation \eqref{eq::accmv2::ipw::est::SA} is identical. 
From a practical perspective, we modify the exponential tilting as $\exp(\delta_{\bar a}^T (\ell_{\bar a} - 7))$. This allows the missingness of HbA1c measurements to 
follow the intuition that a healthier patient with controlled HbA1c levels are more likely to miss their HbA1c measurements, while a sick patients are less likely to miss their HbA1c measurement. For example, with a negative $\delta$, $\delta (\ell_{\bar a} - 7)$ is more likely to be positive for  healthier patients (because their HbA1c levels are more likely to be less than 7) and thus increase the missing probability. Similarly, a negative $\delta$ will decrease the missing probability for sicker patients. For completeness, we also displays the results with a positive $\delta$, which will reverse the relationship between health status and HbA1c missing probability. 

Figure \ref{fig::sa_summary_measures} show the estimates and 95\% confidence intervals for $\E[Y_4]$, $P(Y_3 \leq 7, Y_4 \leq 7)$ and $\E[Y_3 + Y_4] / 2$ as we vary the sensitivity parameter $\delta$. We can see that the results highly agree with our intuition. When $\delta$ is negative and the magnitude of $\delta$ increases, the mean HbA1c levels decreases as we take into account of the fact that healthier patients with lower HbA1c values are more  likely to be missing. For the same reason, the proportion of patients having their HbA1c levels controlled also increases. In the unrealistic scenario that $\delta$ is positive, we observe opposite results.

\subsection{Marginal Parametric model}

\begin{table}[h]
  \caption{Linear regression results for the diabetes data set: $\E[Y_4 | Y_2, Y_3] = \beta_0 + \beta_1 Y_2 + \beta_2 Y_3$.}
  \begin{center}
  \begin{tabular}{ c c c c }
  \hline
  \hline
   Methods   &  $\beta_0$  & $\beta_1$ & $\beta_2$ \\
   \hline
  IPW & 1.120 (0.807 - 1.438) & 0.148 (0.059 - 0.230) & 0.690(0.590 - 0.793) \\
  Complete Case & 1.364 (1.127 - 1.601) & 0.104 (0.048 - 0.160) & 0.701(0.645 - 0.758) \\
  \hline
  \hline
  \end{tabular}
  \end{center}
  \label{table::real_data_results_regression}
\end{table}

\begin{figure}
  \centering
  \includegraphics[width=\textwidth]{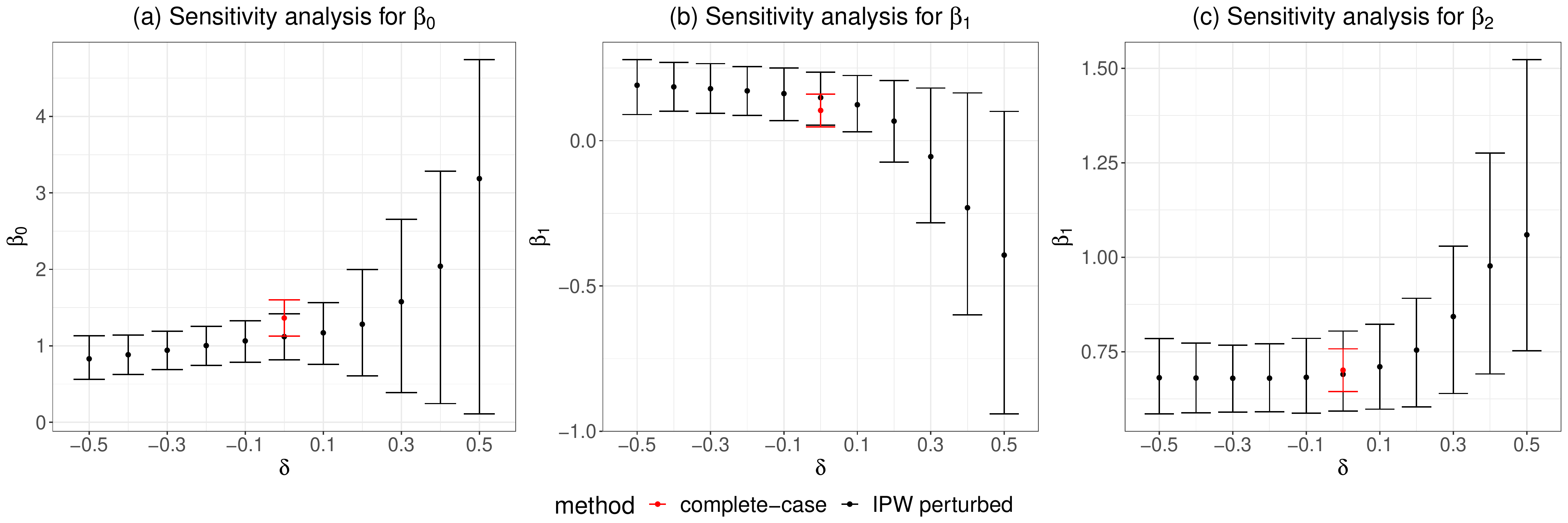}
  \caption{Sensitivity analysis of the ACCMV assumption by exponential tilting. We examine how the parameter estimates and confidence intervals changes with respect to different values of the sensitivity parameter $\delta$.}
  \label{fig::sa_marginal_parametric}
\end{figure}

Further, we also want to study the linear relationship between $Y_2, Y_3$ and $Y_4$. Our intuition is to predict $Y_4$, $Y_3$ should be more important compared to $Y_2$. We consider estimating the linear regression model as follows:
\begin{align*}
  \E[Y_4 | Y_2, Y_3] = \beta_0 + \beta_1 Y_2 + \beta_2 Y_3
\end{align*}
For the linear regression model, the primary variable is $L = (Y_2, Y_3, Y_4)$ and the auxillary variables are $X = (Y_0, Y_1)$. 
Table \ref{table::real_data_results_regression} shows that indeed both $Y_2$ and $Y_3$ has a positive association with $Y_4$ and $Y_3$ has a stronger association than $Y_2$. Figure \ref{fig::sa_marginal_parametric} shows the estimates and 95\% confidence intervals for $\beta_0, \beta_1, \beta_2$ as $\delta$ varies. We can see that when $\delta$ is negative, the estimates are quite robust and do not change much. 


\section{Conclusion}
\label{sec:conc}
In this paper, we introduced the ACCMV assumption to handle nonmonotone and MNAR data. Our ACCMV assumption allows a much larger set of observations to be used for identification compared to the traditional CCMV assumption. Thus, ACCMV is particularly suitable for analyzing data sets with few complete cases. 
We further proposed IPW, regression adjustment and multiply-robust estimators. We also studied their asymptotic and efficiency theories.  We then proposed a sensitivity analysis approach for the ACCMV model for the IPW estimator. Our simulation studies confirm the validity of the assumption.  The real data results also highlight the effect of missing data on the final estimate and the importance of efficiently handling the missing data. 

So far, we have focused on the first-year's data for the HbA1c measurements. However, the diabetes patients are followed up to 11 years and a patient could potentially have up to 44 measurements. Thus, it is helpful to also consider longer history of HbA1c measurements as this can provide more information of a patient. In particular, it will be of interest to recover the whole trajectory for a patient who has a bunch of missing values. This is a much more challenging task as the missing patterns increase exponentially when the number of measurements increases. 
We leave the trajectory recovery problem to future work. 

\acks{
We would like to acknowledge support for this project
 from the National Science Foundation (NSF grant DMS 195278 and 2112907 and CAREER award DMS 2141808)
 and the National Institute of Health (NIH grant U24 AG072122).
}


\newpage

\appendix
\gdef\thesection{Appendix \Alph{section}}

\section{Proof for Single Primary Variable} \label{sec:proof}

\begin{proof}[ of Proposition \ref{prop::NP}]
$p(\ell, A = 1)$ is clearly identifiable. 
 We can write $p(\ell, A = 0)$ as 
 \begin{align*}
  p(\ell, A = 0) = \sum_r p(\ell, R = r, A = 0)
 \end{align*}
 and we further have 
 \begin{align*}
   p(\ell, R = r, A = 0) & = \int p(\ell, x_r, R = r, A  = 0) d x_r = \int p(\ell | x_r, R = r, A = 0) p(x_r, R = r, A = 0) d x_r \\
   & = \int p(\ell | x_r, R \geq r, A = 1) p(x_r, R = r, A = 0) d x_r
 \end{align*}
 Thus, $p(\ell, R = r, A = 0)$ is identifiable under ACCMV, which implies that $p(\ell, A = 0)$ is identifiable. 
\end{proof}

\begin{proof}[ Of Lemma \ref{lemma::accmv_ipw}]
We have 
\begin{align*}
   & p(\ell | x_r, R = r, A = 0) = p(\ell | x_r, R \geq r, A = 1) \\
   & \Leftrightarrow \frac{p(\ell , x_r, R = r, A = 0)}{p(x_r, R = r, A = 0)} = \frac{p(\ell, x_r, R \geq r, A = 1)}{p(x_r, R \geq r, A = 1)} \\
   & \Leftrightarrow \frac{p(R = r, A = 0, \ell, x_r)}{p(R \geq r, A = 1, \ell, x_r)} = \frac{p(R = r, A = 0, x_r)}{p(R \geq r, A = 1, x_r)} \\
   & \Leftrightarrow \frac{p(R = r, A = 0 | \ell, x_r)}{p(R \geq r, A = 1 | \ell, x_r)} = \frac{p(R = r, A = 0 | x_r)}{p(R \geq r, A = 1 | x_r)}
\end{align*}
\end{proof}

We start by giving the asymptotic linear expansion of $\hat \alpha_r$. 
Given 
$$
O_r(X_r) = \frac{P(R = r, A = 0 | X_r)}{P(R \geq r, A = 1 | X_r)} = O_r(X_r; \alpha_r)
$$
we have $P(R = r, A = 0 | X_r,  \{R \geq r, A = 1\} \cup \{R = r, A = 0\}; \alpha_r) = \frac{O_r(X_r; \alpha_r)}{1 + O_r(X_r;  \alpha_r)}$. The log-likelihood has the following form:
\begin{align*}
l_n(\alpha_r) & = \frac{1}{n} \sum_{i=1}^n [I(R_i = r, A_i = 0)\log O_r(X_{i,r}; \alpha_r) - \{I(R_i = r, A_i = 0) + \\
& I(R_i \geq r, A_i = 1)\} \log(1 + O_r(X_{i, r}; \alpha_r))], 
\end{align*}
The score is 
\begin{align*}
S_n(\alpha_r) &= \frac{1}{n} \sum_{i=1}^n \left[ 
I(R_i = r, A_i = 0) \frac{\nabla_{\alpha_r} O_r(X_{i, r}; \alpha_r)}{O_r(X_{i, r}; \alpha_r)} \right.\\
& \left. - \{I(R_i = r, A_i = 0) + I(R_i \geq r, A_i = 1)\} \frac{\nabla_{\alpha_r} O_r(X_{i, r}; \alpha_r)}{1 + O_r(X_{i, r}; \alpha_r)}
\right]
\end{align*}
Further, consider $S(\alpha_r)$ such that $S_n(\alpha_r) \rightarrow_p S(\alpha_r)$ and assuming that $\alpha_r^*$ is the unique solutions for $S(\alpha_r) = 0$. Next, we assume that $\sup_{\alpha_r} \| S_n(\alpha_r) - S_n(\alpha_r)\| = o_P(1)$. Then by theorem 5.9 of \cite{van2000asymptotic}, we have that $\hat{\alpha}_r \rightarrow_p \alpha_r^*$. 
Further,  we have that 
\begin{align*}
& \nabla S_n(\alpha_r) = \frac{1}{n} \sum_{i=1}^n \left[ I(R_i = r, A_i = 0) 
\left(  \frac{\nabla^2_{\alpha_r} O_r(X_{i, r}; \alpha_r)}{O_r(X_{i, r}; \alpha_r)} - \frac{\nabla_{\alpha_r} O_r(X_{i, r}; \alpha_r)^{\otimes 2}}{O_r^2(X_{i, r}; \alpha_r)}
\right) \right. \\
& \left. 
- \{I(R_i = r, A_i = 0) + I(R_i \geq r, A_i = 1)\} \left( 
\frac{\nabla^2_{\alpha_r} O_r(X_{i, r}; \alpha_r) (1 + O_r(X_{i, r}; \alpha_r)) - \nabla_{\alpha_r} O_r(X_{i, r}; \alpha_r)^{\otimes 2}}{(1 + O_r(X_{i, r}; \alpha_r))^2}
\right)
\right]
\end{align*}
and that $\nabla S_n(\alpha_r) \rightarrow_p -\Sigma(\alpha_r)$ with 
\begin{align*}
& \Sigma(\alpha_r^*)  = \E \left[
 P(R \geq r, A = 1 | X_r) \frac{\nabla^2_{\alpha_r} O_r(X_{r}; \alpha_r^*) (1 + O_r(X_{r}; \alpha_r^*)) - \nabla_{\alpha_r} O_r(X_{r}; \alpha_r^*)^{\otimes 2}}{1 + O_r(X_{r}; \alpha_r^*)} \right. \\
 & \left. - 
 P(R \geq r, A = 1 | X_r) \frac{\nabla^2_{\alpha_r} O_r(X_{r}; \alpha_r^*) O_r(X_r; \alpha_r^*) - \nabla_{\alpha_r} O_r(X_{r}; \alpha_r^*)^{\otimes 2}}{O_r(X_{r}; \alpha_r^*)} 
\right]  \\
& = \E \left[ P(R \geq r, A = 1 | X_r) \nabla_{\alpha_r} O_r(X_r; \alpha_r)^{\otimes 2} \frac{1}{O_r(X_r; \alpha_r^*)(1 + O_r(X_r; \alpha_r^*))}
\right]
\end{align*}
Under appropriate assumptions (see theorem 5.21 of \cite{van2000asymptotic}), we have that 
\begin{align*}
\sqrt{n}(\hat{\alpha}_r - \alpha_r^*) = \frac{1}{\sqrt{n}} \sum_{i=1}^n \psi_{r, \alpha_r^*}(X_{i, r}, R_i, A_i) + o_p(1) \rightarrow_d N(0, \Sigma(\alpha_r^*)^{-1})
\end{align*}
where 
\begin{align*}
\psi_{r, \alpha_r}(X_r, R, A) & = \Sigma(\alpha_r)^{-1} \left[ I(R = r, A = 0) \frac{\nabla_{\alpha_r} O_r(X_r; \alpha_r)}{O_r(X_r; \alpha_r)} - \right. \\
& \left. \{I(R = r, A = 0) + I(R \geq r, A = 1)\} \frac{\nabla_{\alpha_r} O_r(X_r; \alpha_r)}{1 + O_r(X_r; \alpha_r)}\right]
\end{align*}

\begin{proof}[ of \autoref{thm::thm_ipw_single}] We prove the asymptotic normality of $\hat \theta_{\sf IPW}$ through its asymptotic linear form. 
For notational convenience, we denote $g(L, X, R, A; \alpha) =  f(L)I(A = 1)[1 + \sum_r O_r(X_r; \alpha_r) I(R \geq r)]$.  Then, we have that $\hat \theta_{\sf IPW} = \frac{1}{n} \sum_{i=1}^n  g(L_i, X_i, R_i, A_i; \hat \alpha)$.  We can rewrite $\hat \theta_{\sf IPW} - \theta_0$ as 
\begin{align*}
\hat \theta_{\sf IPW} - \theta_0 = \underbrace{\frac{1}{n} \sum_{i=1}^n g(L_i, X_i, R_i, A_i; \hat \alpha) - \frac{1}{n} \sum_{i=1}^n  g(L_i, X_i, R_i, A_i; \alpha^*)}_{\mathbf{I}} + \underbrace{\frac{1}{n} \sum_{i=1}^n g(L_i, X_i, R_i, A_i; \alpha^*) - \theta_0}_{\mathbf{II}}
\end{align*}
Term $\mathbf{II}$ is already in the linear expansion form.  For term $\mathbf{I}$, we have that  
\begin{align*}
\sqrt{n} \mathbf{I} & = \frac{1}{\sqrt{n}} \sum_{i=1}^n f(L_i) I(A_i = 1) \sum_r I(R_i \geq r)[ O_r(X_{i, r}; \hat \alpha_r) - O_r(X_{i, r}; \alpha_r^*)] \\
& =  \frac{1}{n} \sum_{i=1}^n f(L_i) I(A_i = 1) \sum_r I(R_i \geq r) \nabla_{\alpha_r} O_r(X_{i, r}; \alpha_r^*)^T  \sqrt{n}(\hat \alpha_r - \alpha_r^*) + o_p(1) \\
& = \sum_r \frac{1}{n} \sum_{i=1}^n f(L_i) I(A_i = 1) I(R_i \geq r) \nabla_{\alpha_r} O_r(X_{i, r}; \alpha_r^*)^T \sqrt{n} (\hat \alpha_r - \alpha_r^*) + o_p(1) \\ 
& = \sum_r  \left( \E [ f(L) I(A = 1) I(R \geq r) \nabla_{\alpha_r} O_r(X_r; \alpha_r^*)^T ] + o_p(1) \right) \sqrt{n} (\hat \alpha_r - \alpha_r^*) + o_p(1) \\
& = \sum_r \E[ \nabla_{\alpha_r} g(L, X, R, A, \alpha^*)^T] \frac{1}{\sqrt{n}} \sum_{i=1}^n \psi_{r, \alpha_r^*}(X_{i, r}, R_i, A_i) + o_p(1) \\
& = \frac{1}{\sqrt{n}} \sum_{i=1}^n \sum_r \E [\nabla_{\alpha_r} g(L, X, R, A, \alpha^*)^T] \psi_{r, \alpha_r^*}(X_{i, r}, R_i, A_i) + o_p(1)
\end{align*}
Thus, combined with term $\mathbf{II}$, we have 
\begin{align*}
\sqrt{n}(\hat \theta_{\sf IPW} - \theta_0) = \frac{1}{\sqrt{n}} \sum_{i=1}^n 
\left[
\sum_r \E [\nabla_{\alpha_r} g(L, X, R, A, \alpha^*)^T] \psi_{r, \alpha_r^*}(X_{i, r}, R_i, A_i) + g(L_i, X_i, R_i, A_i; \alpha^*) - \theta_0
\right] + o_P(1)
\end{align*}
and denote $\phi(X, L, R, A; \alpha^*) = \sum_r \E [\nabla_{\alpha_r} g(L, X, R, A, \alpha^*)^T] \psi_{r, \alpha_r^*}(X_{r}, R, A) + g(L, X, R, A; \alpha^*) - \theta_0$.  Then we have $\sqrt{n}(\hat \theta_{\sf IPW} - \theta_0) \rightarrow_d N(0, \sigma^2_{\sf IPW})$ with $\sigma^2_{\sf IPW} = \Var[\phi(X, L, R, A; \alpha^*)]$. 
\end{proof}

Similarly,  we first give the asymptotic linear expansion of $\hat \beta_r$.  The estimating equation for $\beta_r$ now has the following form:
\begin{align*}
S_n(\beta_r) = \frac{1}{n}\sum_{i=1}^n I(R_i \geq r, A_i = 1)(m_{r, 0}(X_{i, r}; \beta_r) - f(L_i)) \nabla_{\beta_r} m_{r, 0}(X_{i, r}; \beta_r)
\end{align*}
Again for parametric models, under appropriate assumptions (see theorem 5.9 of \cite{van2000asymptotic}), we have that $\hat \beta_r \rightarrow_p \beta_r^*$.  Further, we have that 
\begin{align*}
\nabla S_n(\beta_r) = \frac{1}{n} \sum_{i=1}^n I(R_i \geq r, A_i = 1)[\nabla^2_{\beta_r} m_{r, 0}(X_{i, r}; \beta_r) (m_{r, 0}(X_{i, r}; \beta_r) - f(L_i)) + \nabla_{\beta_r} m_{r, 0}(X_{i, r}; \beta_r)^{\otimes 2}]
\end{align*}
and that $\nabla S_n(\beta_r) \rightarrow_p \nabla S(\beta_r)$ with 
\begin{align*}
\nabla S(\beta_r^*) = \E \left[ I(R \geq r, A = 1) \nabla_{\beta_r} m_{r, 0}(X_r; \beta_r^*)^{\otimes 2} 
\right]
\end{align*}
Next, we have that 
\begin{align*}
\sqrt{n}(\hat \beta_r - \beta_r^*) = \frac{1}{\sqrt{n}} \nabla S(\beta_r^*)^{-1} \sum_{i=1}^n I(R_i \geq r, A_i = 1) (f(L_i) - m_{r, 0}(X_{i, r}; \beta_r^*)) \nabla_{\beta_r} m_{r, 0}(X_{i, r}; \beta_r^*) + o_p(1)
\end{align*}
with $\psi_{r, \beta_r^*}(L, X_r, R, A) = \nabla S(\beta_r^*)^{-1}I(R \geq r,  A = 1)(f(L) - m_{r, 0}(X_r; \beta_r^*)) \nabla_{\beta_r} m_{r, 0}(X_r; \beta_r^*)$.

\begin{proof}[ of \autoref{thm::thm_ra_single}]
Now we give the proof for the regression adjustment estimation. The proof is very similar to proof of \autoref{thm::thm_ipw_single}. 
  Similarly, denote $h(L, X, R, A; \beta) = f(L) A + \sum_r m_{r, 0}(X_r; \beta_r)(1 - A) I(R = r)$, we have that 
\begin{align*}
\hat \theta_{\sf RA} - \theta = \underbrace{\frac{1}{n} \sum_{i=1}^n h(L_i, X_i, R_i, A_i; \hat \beta) - \frac{1}{n} \sum_{i=1}^n h(L_i, X_i, R_i, A_i; \beta^*)}_{\mathbf{I}} + \underbrace{\frac{1}{n} \sum_{i=1}^n h(L_i, X_i, R_i, A_i; \beta^*) - \theta}_{\mathbf{II}}
\end{align*}
For term $\mathbf{I}$, we have 
\begin{align*}
\sqrt{n} \mathbf{I} & =  \frac{1}{\sqrt{n}} \sum_{i=1}^n \sum_r \left[ m_{r, 0}(X_{i, r}; \hat \beta_r) - m_{r, 0}(X_{i, r}; \beta_r^*)\right] I(R_i = r, A_i = 0) \\
& = \sum_r \E[ \nabla_{\beta_r} m_{r, 0}(X_r; \beta_r^*)^T I(R = r, A = 0)] \sqrt{n} (\hat{\beta}_r - \beta_r^*) + o_p(1) \\
& = \frac{1}{\sqrt{n}} \sum_{i=1}^n \sum_r \E[\nabla_{\beta_r} m_{r, 0}(X_r; \beta_r^*)^T I(R = r, A = 0)] \psi_{r, \beta_r^*}(L_i, X_{i, r}, R_i, A_i) + o_p(1)
\end{align*}
Thus, combined with term $\mathbf{II}$, denote 
$$
\phi(L, X, R, A; \beta^*) = \sum_r \E[\nabla_{\beta_r} m_{r, 0}(X_r; \beta_r^*)^T I(R = r, A = 0)] \psi_{r, \beta_r^*}(L, X_{r}, R, A) + h(L, X, R, A; \beta^*) - \theta
$$
we have
\begin{align*}
\sqrt{n}(\hat \theta_{\sf RA} - \theta) = \frac{1}{\sqrt{n}} \sum_{i=1}^n \phi(L_i, X_i, R_i, A_i; \beta^*) + o_p(1) \rightarrow_d N(0, \sigma_{\sf RA}^2)
\end{align*}
with $\sigma^2_{\sf RA} = \Var[\phi(X, L, R, A)]$. 
\end{proof}

\section{Proof of Multiply-robustness for Single Variable} \label{sec::mr_proof}

\begin{proof}[ of Theorem~\ref{thm::accmv1::eff}]
Recall that the IPW formulation for $\theta_{0, r}$ is 
\begin{align*}
\theta_{0, r} =  \E[ f(L) I(A = 1) I(R \geq r) O_r(X_r; \alpha_r^*)] 
= \int f(\ell) O_r(x_r) I(s \geq r) I(a = 1) p_0(\ell, x_r, s, a) d\ell d x_r ds da.
\end{align*}
and $p_0(\ell,  x_r,  s,  a)$ is the true model.  We consider a pathwise perturbation $p_\epsilon(\ell,  x_r,  s,  a) = p_0(\ell,  x_r,  s, a) (1 + \epsilon \cdot g(\ell,  x_r,  s, a))$ such that $g$ satisfies 
$$
\int p_0(\ell,  x_r,  s,  a) g(\ell,  x_r, s, a) d \ell dx_r  ds da = 0
$$ 
Under $p_\epsilon$,  denote $\theta_{0, r}$ as $\theta_{0, r}^\epsilon$.  We derive the EIF using the semi-parametric theory (see section 25.3 of \cite{van2000asymptotic}),  the EIF is a function ${\sf EIF}(\ell,  x_r,  s,  a)$ such that $\E[{\sf EIF}(L,  X_r,  R,  A)] = 0$ and 
\begin{align*}
\lim_{\epsilon \rightarrow 0} \frac{\theta_{0, r}^\epsilon - \theta_{0, r}}{\epsilon} = \int  {\sf EIF}(\ell,  x_r,  s,  a)  p_0(\ell,  x_r,  s,  a) g(\ell,  x_r, s, a) d \ell dx_r  ds da
\end{align*}
Under model $p_\epsilon$,  we also have perturbed odds $O_{r, \epsilon}(x_r)$.  We denote $\Delta O_r(x_r) = O_{r, \epsilon}(x_r) - O_{r, 0}(x_r)$.   Then,  a direct computation shows that 
\begin{align*}
\theta_{0, r}^\epsilon & = \int f(\ell) O_{r, \epsilon}(x_r) I(s \geq r) I(a = 1) p_\epsilon(\ell,  x_r,  s, a)  d \ell dx_r  ds da  \\
& = \theta_{0, r} + \epsilon \underbrace{\int f(\ell)  O_{r, 0}(x_r) I(s \geq r) I(a = 1) p_0(\ell,  x_r,  s, a) g(\ell,  x_r,   s,  a) d \ell dx_r ds d a}_{\mathbf{A}} \\
& + \underbrace{\int f(\ell)  \Delta O_r(x_r) I(s \geq r) I(a = 1) p_0(\ell,  x_r,  s, a) d \ell dx_r ds d a}_{\mathbf{B}} + O(\epsilon^2)
\end{align*}
Part $\mathbf{A}$ is already in the form of an EIF.  For part $\mathbf{B}$,  we need to derive $\Delta O_r(x_r)$.  
Now we expand the difference $\Delta O_r(x_r)$ as following:
\begin{align*}
\Delta O_r(x_r) & = O_{r, \epsilon}(x_r) - O_{r, 0}(x_r) = \frac{p_{\epsilon}(R = r,  A = 0,  x_r)}{p_\epsilon(R \geq r,  A = 1,  x_r)} - \frac{p_0(R = r,  A = 0,  x_r)}{p_0(R \geq r, A = 1,  x_r)} \\
& = \frac{1}{p_0(R \geq r,  A = 1,  x_r)} \left[  \Delta p(R = r,  A = 0,  x_r) - O_{r, 0}(x_r) \Delta p(R \geq r, A = 1, x_r) \right] + O(\epsilon^2)
\end{align*}
with
\begin{align*}
\Delta p(R = r,  A = 0,  x_r) & = p_\epsilon(R = r, A = 0, x_r) - p_0(R = r,  A = 0, x_r) \\
& = \epsilon \int I(s = r)I(a = 0) p_0(\ell,  x_r,  s, a) g(\ell,  x_r,   s,  a) d \ell ds d a \\
\Delta p(R \geq r,  A = 1, x_r) & = p_\epsilon(R \geq r, A = 1, x_r) - p_0(R \geq r,  A = 1, x_r) \\
& = \epsilon \int I(s \geq r) I(a = 1) p_0(\ell,  x_r,  s, a) g(\ell,  x_r,   s,  a) d \ell ds d a
\end{align*}
Thus,  the difference $\Delta O_r(x_r)$ can be rewritten as
\begin{align*}
\Delta O_r(x_r) = & \frac{\epsilon}{p_0(R \geq r, A = 1, x_r)} \int \left[ I(s = r)I(a = 0) - O_{r, 0}(x_r) I(s \geq r) I(a = 1) \right] \\
& \times p_0(\ell,  x_r,  s, a) g(\ell,  x_r,   s,  a) d \ell ds d a + O(\epsilon^2)
\end{align*}
Now part $\mathbf{B}$ can be rewritten as 
\begin{align*}
\mathbf{B} & = \int  \Delta O_r(x_r) f(\ell) I(s \geq r) I(a = 1) p_0(\ell,  x_r,  s, a) d \ell dx_r ds d a \\
& =  \int \Delta O_r (x_r) p_0(R \geq r,  A = 1, x_r)  f(\ell) p_0(\ell | R \geq r,  A = 1, x_r) d \ell d x_r \\
& = \int \Delta O_r (x_r) p_0(R \geq r,  A = 1, x_r)  \underbrace{\left\{ \int f(\ell) p_0(\ell | R \geq r,  A = 1, x_r) d \ell  \right\}}_{\E [f(L) | R \geq r, A = 1,  X_r = x_r] = m_{r, 0}(X_r)}  dx_r \\
& = \int \Delta O_r(x_r)  p_0(R \geq r, A = 1, x_r) m_{r, 0}(x_r)  dx_r \\
& = \epsilon \int [I(s = r)I(a = 0) - O_{r, 0}(x_r) I(s \geq r) I(a = 1)] m_{r, 0}(x_r) p_0(\ell,  x_r,  s, a) g(\ell,  x_r,   s,  a) d \ell d x_r ds d a + O(\epsilon^2)
\end{align*}
Thus,  combining part $\mathbf{A}$ and $\mathbf{B}$,  we conclude that 
\begin{align*}
{\sf EIF}_{r, 0}(\ell,  x_r,  s, a) & = f(\ell) O_{r, 0}(x_r) I(s \geq r) I(a = 1)\\
& + [I(s = r) I(a = 0) - O_{r, 0}(x_r) I(s \geq r) I(a = 1)] m_{r, 0}(x_r) - \theta_{0, r} \\
& = [f(\ell) - m_{r, 0}(x_r)] O_r(x_r) I(s \geq r, a = 1) + I(s = r, a = 0) m_{r, 0}(x_r) - \theta_{0, r}
\end{align*}
\end{proof}

We first discuss assumptions for the Donsker classes ${\cal F}_r$ and ${\cal G}_r$ where we have assumed that $\hat O_r \in {\cal F}_r$ and $\hat m_{r, 0} \in {\cal G}_r$. We denote the $L_2(Q)$ norm as 
$$
\|f\|_{Q, 2} = \int f^2 dQ
$$
for a probability measure $Q$. We assume that ${\cal F}_r$ satisifies the following uniform entropy condition \citep{van1996weak}: 
\begin{align*}
  \int_0^{\infty} \sup_Q \sqrt{\log N(\epsilon \|F_r\|_{Q, 2}, {\cal F_r}, L_2(Q))} d \epsilon < \infty
\end{align*}
where the supremum of $Q$ is taken over all finitely discrete probability measures on $X$. $N(\epsilon \|F_r\|_{Q, 2}, {\cal F_r}, L_2(Q))$ is the covering number of class ${\cal F}_r$ with respect to the $L_2(Q)$ norm (see Definition 2.1.5 of \cite{van1996weak}) and $F_r$ is an envelope function of ${\cal F}_r$ such that $\P F_r^2 < \infty$ with $P$ being the probability measure for $X$. Similarly we can assume that ${\cal G}_r$ satisfy a same uniform entropy condition with envelop function $G_r$. We also assume that ${\cal F}_r$ and ${\cal G}_r$ are suitably measurable (see Definition 2.3.3 of \cite{van1996weak}) and $\P F_r^2 G_r^2 < \infty$.  Further,  we let $\P_n f = \frac{1}{n} \sum_{i=1}^n f(X_i)$ and $\P_0 f = \int f(x) dP(x)$. 
\begin{proof}[ of Theorem~\ref{thm::mr1}]
By assumption (M1),  we have  $\| \hat O_r - O_r^*\|_{L_2(P)} = o_P(1)$ and $\| \hat m_{r, 0} - m_{r, 0}^*\|_{L_2(P)} = o_p(1)$.  The true functions are denoted as $O_r(x_r)$ and $m_{r, 0}(x_r)$.  When the models are correct, we have $O_r(x_r) = O_r^*(x_r)$ and $m_{r, 0}(x_r) = m_{r, 0}^*(x_r)$. 
The multiply-robust estimator has the following form:
\begin{align*}
\hat \theta_{\sf MR} & = \frac{1}{n} \sum_{i=1}^n [ f(L_i) I(A_i = 1)  + \\
&  +  \sum_r \left( \{f(L_i)-\hat m_{r,0}(X_{i,r}) \} \hat O_r(X_{i,r}) I(R_i\geq r)I(A_i=1) + \hat m_{r,0}(X_{i,r}) I(R_i=r)I(A_i=0) \right) ]
\end{align*}

We first consider another estimator that replaces the estimated functions by the true functions in the multiply-robust estimator.  
\begin{align*}
\tilde \theta_{\sf MR} & = \frac{1}{n} \sum_{i=1}^n [ f(L_i) I(A_i = 1)  + \\
&  +  \sum_r \left( \{f(L_i)-  m_{r,0}(X_{i,r})\}  O_r(X_{i,r}) I(R_i\geq r)I(A_i=1) +  m_{r,0}(X_{i,r}) I(R_i=r)I(A_i=0) \right) ]
\end{align*}
It is not hard to prove that $\tilde \theta_{\sf MR} \rightarrow_p \theta$ and 
$$
\sqrt{n}(\tilde \theta_{\sf MR} - \theta) \rightarrow_d N(0, \sigma^2_{\sf eff})
$$
where $\sigma^2_{\sf eff}$ is the efficiency bound for estimating $\theta$. Further we have that 
\begin{equation}
\begin{aligned}
\widehat{\theta}_{\sf MR}  & = \widetilde{\theta}_{\sf MR} + \frac{1}{n} \sum_{i=1}^n 
\left\{  \underbrace{\sum_r I(R_i \geq r) I(A_i = 1) (\hat O_r(X_{i, r}) - O_r(X_{i, r}))  (f(L_i) - m_{r, 0}(X_{i, r}))}_{(\mathbf{I})}  \right. \\
& + \underbrace{\sum_r (\hat m_{r, 0}(X_{i, r}) - m_{r, 0}(X_{i, r})) \left[I(R_i = r) I(A_i = 0) - I(R_i \geq r) I(A_i = 1) O_r(X_{i, r}) \right]}_{(\mathbf{II})} \\
& \left. - \underbrace{\sum_r I(R_i \geq r) I(A_i = 1) \left[ \hat m_{r, 0}(X_{i, r}) - m_{r, 0}(X_{r, i})\right] \left[ \hat O_r(X_{i, r}) - O_r(X_{i, r}) \right]}_{(\mathbf{III})} \right\}
\end{aligned}
\label{eqn::eq_mr}
\end{equation}

We first prove the multiply-robust property when the odds are correctly specified and the regression functions are misspecified. 
Thus, $O_r(x_r) = O_r^*(x_r)$ and $m_{r, 0}(x_r) \neq m_{r, 0}^*(x_r)$. 
Since $\| \hat m_{r, 0} - m_{r, 0} \|_{L_2(P)} \| \hat O_r - O_r\|_{L_2(P)} = o_P(1)$, we have $\|\hat O_r - O_r\|_{L_2(P)} = o_P(1)$. 
Denote $g_{r, n}(x_r, l, s, a) = I(s \geq r) I(a = 1)(\hat O_r(x_r) - O_r(x_r))(f(l) - m_{r, 0}(x_r))$ and similarly 
$$
g_{r, 0}(x_r, l, s, a) = I(s \geq r) I(a = 1) (O_r^*(x_r) - O_r(x_r))(f(l) - m_{r, 0}(x_r)) = 0.
$$  Then we can write term $\mathbf{I}$ in \eqref{eqn::eq_mr} for pattern $r$ as
\begin{align*}
\frac{1}{n} \sum_{i=1}^n I(R_i \geq r) I(A_i = 1) (\hat O_r(X_{i, r}) - O_r(X_{i, r}))(f(L_i) - m_{r, 0}(X_{i, r})) = \P_n g_{r, n} = \P_n (g_{r, n}  -g_{r, 0})
\end{align*}
Given that $\|\hat O_r - O_r\|_{L_2(P)} = o_P(1)$, we have $\| g_{r, n} - g_{r, 0}\|_{L_2(P)} = o_P(1)$ as $f(l), m_{r, 0}(x_r)$ are uniformly bounded. 
Further, 
we have  
\begin{align*}
g_{r, n}(x_r, l, s, a) &= I(s \geq r)I(a = 1) (\hat O_r(x_r) - O_r(x_r))(f(l) - m_{r, 0}(x_r)) \\
& = I(s \geq r)I(a = 1)(f(l) - m_{r, 0}(x_r)) \hat O_r(x_r) - g_{1, r}(x_r, l, s, a)
\end{align*}
where $g_{1, r}(x_r, l, s, a)$ is a function that does not involve $\hat O_r$. 
As $\hat O_r(x_r)$ is in a Donsker class ${\cal F}_r$, we have $g_{r, n}$ is also in a Donsker class 
$$
{\cal F}^* = \{I(s \geq r) I(a = 1)(f(l) - m_{r, 0}(x_r)) h(x_r) - g_{1, r}(x_r, l, s, a): h \in {\cal F} \}
$$
by Example 2.10.7 and 2.10.23 of \cite{van1996weak} and the assumptions we made before this proof. 
By Lemma 19.24 of \cite{van2000asymptotic}, we then have 
$$
(\P_n - \P_0)(g_{r, n} - g_{r, 0}) = o_P(n^{-1/2}) \Rightarrow (\P_n - \P_0) g_{r, n} = o_P(n^{-1/2}) \Rightarrow \P_n g_{r, n} = o_P(n^{-1/2})
$$
realizing that $\P_0 g_{r, n} = 0$. For term $\mathbf{II}$, similarly define
$$
g'_{r, n}(x_r, s, a) = (\hat m_{r, 0}(x_r) - m_{r, 0}(x_r))[ I(s = r) I(a = 0) - I(s \geq r) I(a = 1) O_r(x_r)]
$$
and 
$$
g'_{r, 0}(x_r, s, a) = (m_{r, 0}^*(x_r) - m_{r, 0}(x_r))[I(s = r) I(a = 0) - I(s \geq r) I(a = 1) O_r(x_r)]
$$
Then we can rewrite term $\mathbf{II}$ in \eqref{eqn::eq_mr} for pattern $r$ as
\begin{align*}
  & \frac{1}{n} \sum_i (\hat m_{r, 0}(X_{i, r}) - m_{r, 0}(X_{i, r}))[I(R_i = r)I(A_i = 0) - I(R_i \geq r)I(A_i = 1)O_r(X_{i, r})] \\
  & = (\P_n - \P_0)(g'_{r, n} - g'_{r, 0})  \\
  & + \frac{1}{n} \sum_{i=1}^n (m_{r, 0}^*(X_{i, r}) - m_{r, 0}(X_{i, r}))[I(R_i = r)I(A_i = 0) - I(R_i \geq r)I(A_i = 1)O_{r}(X_{i, r})]
\end{align*}
First note that 
\begin{align*}
  \frac{1}{n} \sum_{i=1}^n (m_{r, 0}^*(X_{i, r}) - m_{r, 0}(X_{i, r}))[I(R_i = r)I(A_i = 0) - I(R_i \geq r)I(A_i = 1)O_{r}(X_{i, r})] \rightarrow_p 0
\end{align*}
This implies that term $\mathbf{II}$ in \eqref{eqn::eq_mr} for pattern $r$ can be written as 
\begin{align*}
  (\P_n - \P_0)(g'_{r, n} - g'_{r, 0}) + o_P(1)
\end{align*}
Then again given that $\hat m_{r, 0}(x_r)$ is in a Donsker class ${\cal G}_r$ and $\| \hat m_{r, 0} - m_{r, 0}^*\|_{L_2(P)} = o_P(1)$, we have that
$\|g'_{r, n} - g'_{r, 0}\|_{L_2(P)} = o_P(1)$ and $g'_{r, n}$ is also in a Donsker class by a similar reasoning as before. By Lemma 19.24 of \cite{van2000asymptotic}, we have 
$$
(\P_n - \P_0)(g'_{r, n} - g'_{r, 0}) = o_P(n^{-1/2}) \Rightarrow (\P_n - \P_0)g'_{r, n} = (\P_n - \P_0) g'_{r, 0} + o_P(n^{-1/2}) 
$$
This implies that 
\begin{align*}
\P_n g'_{r, n}  = o_P(1) + o_P(n^{-1/2}) + \P_0 g'_{r, n} = o_P(1) + o_P(n^{-1/2}) = o_P(1)
\end{align*}
as $\P_0 g'_{r, n} = 0$. For term $\mathbf{III}$, we can similarly define 
$$
h_n(x_r, s, a) = (\hat m_{r, 0}(x_r) - m_{r, 0}(x_r))(\hat O_r(x_r) - O_r(x_r)) I(s \geq r) I(a = 1)
$$
and $h_0(x_r, s, a) = 0$.
Then $h_n(x_r, s, a)$ is in a Donsker Class by Example 2.10.23 of \cite{van1996weak}.  Given that $\|\hat O_r - O_r\|_{L_2(P)} = o_P(1)$ and $\hat m_{r, 0}, m_{r, 0}$ are uniformly bounded, we have $\| h_n  - h_0 \|_{L_2(P)} = o_P(1)$.  Then by Lemma 19.24 of \cite{van2000asymptotic}, we have 
\begin{align*}
(\P_n - \P_0)(h_n - h_0) = o_P(n^{-1/2}) \Rightarrow (\P_n - \P_0) h_n = o_P(n^{-1/2}) \Rightarrow \P_n h_n = \P_0 h_n + o_P(n^{-1/2})
\end{align*}
and 
\begin{align*}
\P_0 h_n \leq \| \hat m_{r, 0} - m_{r, 0}\|_{L_2(P)} \| \hat O_r - O_r \|_{L_2(P)} = o_P(1)
\end{align*}
given the assumption that 
$$
\sum_r \| \hat m_{r, 0} - m_{r, 0} \|_{L_2(P)} \|\hat O_r - O_r\|_{L_2(P)} = o_P(1)
$$
Above results imply that  $\hat \theta_{\sf MR} = \tilde \theta_{\sf MR} + o_P(1) \rightarrow_p \theta$. 
By similar reasoning,  when we have odds function misspecified and 
regression function correctly specified, we also have $\hat \theta_{\sf MR} \rightarrow_p \theta$. 

When we have both model correctly specified, by a similar proof as above, term $\mathbf{I}$ has leading term on the order of $o_P(n^{-1/2})$ for each pattern $r$.  Similarly, $\mathbf{II}$ is also $o_P(n^{-1/2})$. 
Thus,  we have
\begin{align*}
\hat \theta_{\sf MR} - \tilde \theta_{\sf MR} = -\frac{1}{n} \sum_{i=1}^n \sum_r I(R_i \geq r) I(A_i = 1)  (\widehat{m}_{r, 0}(X_{i, r}) - m_{r, 0}(X_{r, i}))(\widehat{O}_r(X_{i, r}) - O_r(X_{i, r})) + o_p(n^{-1/2})
\end{align*}
Finally, for term $\mathbf{III}$, by the same proof, we have 
\begin{align*}
\P_n h_n = \P_0 h_n + o_P(n^{-1/2}) \leq \| \hat m_{r, 0} - m_{r, 0}\|_{L_2(P)} \| \hat O_r - O_r \|_{L_2(P)}  + o_P(n^{-1/2}) = o_P(n^{-1/2})
\end{align*}
assuming that 
$$
\sqrt{n} \sum_r \| \hat m_{r, 0} - m_{r, 0} \|_{L_2(P)} \|\hat O_r - O_r\|_{L_2(P)} = o_P(1)
$$

Together,  we have proved that 
\begin{align*}
\sqrt{n} (\hat \theta_{\sf MR} - \theta) = \sqrt{n}(\hat \theta_{\sf MR} - \tilde \theta_{\sf MR}) + \sqrt{n} (\tilde \theta_{\sf MR} - \theta) = o_P(1) + \sqrt{n} (\tilde \theta_{\sf MR} - \theta) \rightarrow_d N(0, \sigma_{\sf eff}^2)
\end{align*}
\end{proof}

In fact, when we use parametric estimators for both $\hat m_{r, 0}$ and $\hat O_r$,  as long as for each pattern $r$, either $m_{r, 0}(x_r; \beta_r^*) = m_{r, 0}(x_{r})$ or $O_{r}(x_r; \alpha_r^*) = O_r(x_r)$, we have the following asymptotic linear expansion for $\hat \theta_{\sf MR}$ as 
\begin{align*}
  &   \sqrt{n} (\hat{\theta}_{\sf MR} - \theta_0) = \frac{1}{\sqrt{n}} \sum_{i=1}^n \left[
    h(L_i, X_i, R_i, A_i; \beta^*, \alpha^*) + f(L_i) I(A_i = 1) + \right.\\
    & \left. 
      \sum_r \left\{\E[\nabla_{\beta_r} h(L, X, R, A; \beta^*)^T] \psi_{r, \beta_r^*}(L, X_r, R, A) + \E[\nabla_{\alpha_r} h(L, X, R, A; \alpha^*)^T] \psi_{r, \alpha_r^*}(X_r, R, A) \right\}  - \theta_0
    \right] + o_P(1)
\end{align*}
with 
\begin{align*}
  h(L, X, R, A; \beta^*, \alpha^*) &= \sum_r\left\{ [f(L) - m_{r, 0}(X_r; \beta_r^*)]O_r(X_r; \alpha_r^*)I(R \geq r)I(A = 1) + \right. \\
  & \left. m_{r, 0}(X_r; \beta_r^*)I(R = r)I(A = 0) \right\}
\end{align*}
This will be used when we run the simulation studies. 

\section{Proof for Multiple Primary Variables} \label{sec::multiple}
Now we present the proof for the results when there are multiple primary variables. The proof for Proposition \ref{prop::NP2} is omitted as it is very similar to the proof in the single variable case. 

First,  for the IPW estimator, we again give the influence function when we estimate $O_{r, a}$ with a parametric model.  We have that 
\begin{align*}
\sqrt{n}(\hat{\alpha}_{r, a} - \alpha_{r, a}^*) = \frac{1}{\sqrt{n}} \sum_{i=1}^n \psi_{r, a}(X_{i, r}, L_{i, a}, R_i, A_i) + o_p(1) \rightarrow_d N(0, \Sigma(\alpha_{r, a}^*)^{-1})
\end{align*}
where 
\begin{align*}
\psi_{r, a}(X_r, L_a, R, A) & = \Sigma(\alpha_{r, a}^*)^{-1} \left[ I(R = r, A = a) \frac{\nabla_{\alpha_{r, a}} O_{r, a}(X_r, L_a; \alpha_{r, a}^*)}{O_{r, a}(X_r, L_a; \alpha_{r, a}^*)} - \right. \\
& \left. \{I(R = r, A = a) + I(R \geq r, A = 1_d)\} \frac{\nabla_{\alpha_{r, a}} O_{r, a}(X_r, L_a; \alpha_{r, a}^*)}{1 + O_{r, a}(X_r, L_a; \alpha_{r, a}^*)}\right]
\end{align*}
with 
\begin{align*}
\Sigma(\alpha_{r, a}^*) = \E \left[ P(R \geq r, A = 1_d | X_r, L_a) \nabla_{\alpha_{r, a}} O_{r, a}(X_r, L_a ; \alpha_{r, a})^{\otimes 2} \frac{1}{O_{r, a}(X_r, L_a; \alpha_{r, a}^*)(1 + O_{r, a}(X_r, L_a; \alpha_{r, a}^*))}
\right]
\end{align*}

\begin{proof}[ of Theorem~\ref{thm::ipw2}] 
The proof is similar to the proof of \autoref{thm::thm_ipw_single} and we directly give the results. Denote 
\begin{align*}
\phi(X, L, R, A, \alpha^*) &= \sum_{r, a \neq 1_d} \left(  \E\left[
f(L) I(R \geq r, A = 1_d) \nabla_{\alpha_{r, a}} O_{r, a}(X_{r}, L_{a}; \alpha_{r, a}^*) 
\right]\psi_{r, a}(X_r, L_a, R, A) \right. \\
& \left. + f(L) I(R \geq r, A = 1_d) O_{r, a}(X_r, L_a) \right) + f(L) I(A = 1_d) - \theta_0
\end{align*}
and we have
\begin{align*}
\sqrt{n} (\hat \theta_{\sf IPW} - \theta_0) = \frac{1}{\sqrt{n}} \sum_{i=1}^n \phi(X_i, L_i, R_i, A_i, \alpha^*) + o_P(1) \rightarrow N(0, \sigma^2_{\sf IPW})
\end{align*}
with $\sigma_{\sf IPW}^2 = \Var[\phi(X, L, R, A; \alpha^*)]$.
\end{proof}

For the regression adjustment method,  we have that 
\begin{align*}
\sqrt{n} (\hat \beta_{r, a} - \beta_{r, a}^*) = \frac{1}{\sqrt{n}} \sum_{i=1}^n \psi_{r, a}(X_{i, r}, L_{i, a}, R_i, A_i) + o_P(1) \rightarrow_d N(0, \Sigma(\beta_{r, a}^*)^{-1} )
\end{align*}
where 
\begin{align*}
\psi_{r, a}(X_{r}, L_{a}, R, A) = \nabla S(\beta_{r, a}^*)^{-1} I(R \geq r, A = 1_d)[ f(L) - m_{r, a}(X_r, L_a; \beta_{r, a}^*)] \nabla_{\beta_{r, a}} m_{r,a}(X_r, L_a, \beta_{r, a}^*)
\end{align*}
with 
\begin{align*}
  \nabla S(\beta_{r, a}^*) = \E[I(R \geq r, A=  1) \nabla_{\beta_{r, a}} m_{r, a}^{\otimes 2}]
\end{align*}

\begin{proof}[ of Theorem~\ref{thm::ra2}]
The proof is again very similar to the proof of \autoref{thm::thm_ra_single} and we directly give the results.  Now we have 
\begin{align*}
\phi(X, L, R, A; \beta^*) &= \sum_{r, a \neq 1_d} \left(
\E[ I(R = r, A = a) \nabla_{\beta_{r, a}} m_{r, a}(X_r, L_a; \beta_{r, a}^*)] \psi_{r, a}(X_r, L_a, R, A) \right. \\
& \left. + m_{r, a}(X_r, L_a; \beta_{r, a}^*) I(R = r, A = a) \right) + f(L) I(A = 1_d)
\end{align*}
Then, we have 
\begin{align*}
\sqrt{n} (\hat \theta_{\sf RA} - \theta_0) = \frac{1}{\sqrt{n}} \sum_{i=1}^n \phi(X_i, L_i, R_i,  A_i; \beta^*) + o_P(1) \rightarrow_d N(0, \sigma_{\sf RA}^2)
\end{align*}
with $\sigma_{\sf RA}^2 = \Var[ \phi(X, L, R, A)]$. 
\end{proof}

We assume that ${\cal F}_{r, a}$ and ${\cal G}_{r, a}$ satisfy the uniform entropy condition with envelop functions $F_{r, a}$ and $G_{r, a}$.  We further assume that ${\cal F}_{r, a}$ and ${\cal G}_{r, a}$ are suitably measurable and $\P F_{r, a}^2 G_{r, a}^2 < \infty$. 
The proof of Theorems \ref{thm::accmv2::eff} and \ref{thm::accmv::mr} is omitted as it is almost identical to the proof of Theorems \ref{thm::accmv1::eff} and \ref{thm::mr1}. 

\section{Proof for Marginal Parametric Model} \label{sec::ap::parametric}
Under mild regularity conditions, we can prove that $\hat \theta \rightarrow_p \theta^*$ by theorem 5.9 of \cite{van2000asymptotic}. 
\begin{proof}[ of theorem \ref{thm::ipw_margin_parametric_model}]
The sample estimating equation for the marginal parametric model under the 
ACCMV assumption is as following:
\begin{align*}
\sum_{i=1}^n s(\hat \theta | L_i) \left[ \sum_{r, a \neq \1_d} O_{r, a}(X_{i, r}, L_{i, a}; \hat \alpha_{r, a}) I(A_i = \1_d) I(R_i \geq r) + I(A_i = \1_d) \right] = 0
\end{align*}
We can define 
$$
\psi_{\theta, \alpha}(L, X, R, A) = s(\theta | L) \left[ \sum_{r, a \neq \1_d} O_{r, a}(X_{r}, L_{a}; \alpha_{r, a}^*) I(A = \1_d) I(R \geq r) + I(A = \1_d) \right]
$$
then we have 
\begin{align*}
\P_n \psi_{\hat \theta, \hat \alpha} = 0  \Rightarrow \P_n \psi_{\hat \theta, \hat \alpha} - \P_n \psi_{\hat \theta, \alpha^*} + \P_n \psi_{ \hat \theta, \alpha^*} = 0 
\end{align*}
Define 
$$
\phi_{\theta, \alpha_{r ,a}}(L, X, R, A) = s(\theta | L)  \nabla_{\alpha_{r, a}} O_{r, a}(X_{r}, L_a; \alpha_{r, a}) I(A = \1_d) I(R \geq r)
$$
Then we have that 
\begin{align*}
\P_n \left[ \psi_{\hat \theta, \hat \alpha}  - \psi_{\hat \theta, \alpha^*} \right] = \sum_{r, a \neq \1_d} \P_n \phi_{\hat \theta, \alpha_{r, a}^*}^T (\hat \alpha_{r, a} - \alpha_{r, a}^*)  + o_P(1) \|\hat \alpha - \alpha\| 
\end{align*}
where we also have 
\begin{align*}
\hat \alpha_{r, a} - \alpha^*_{r, a} = \P_n \xi_{r, a} + o_P(1/\sqrt{n})
\end{align*}
based on our assumption. Thus, put everything together and multiply by $\sqrt{n}$ on both sides of the equation, we have 
\begin{align*}
\sum_{r, a \neq 1_d} \P_n \phi_{\hat \theta, \alpha_{r, a}^*}^T \sqrt{n} \P_n \xi_{r, a} + o_P(1) + \sqrt{n} \P_n \psi_{\hat \theta, \alpha^*} = 0
\end{align*}
next, we have that 
\begin{align*}
\P_n \phi_{\hat \theta, \alpha_{r, a}^*} = \P_n \phi_{\hat \theta, \alpha_{r, a}^*} - \P_0 \phi_{\theta_0, \alpha_{r, a}^*} + \P_0 \phi_{\theta_0, \alpha_{r, a}^*}
\end{align*}
Further, we have 
\begin{align*}
  \P_n \phi_{\hat \theta, \alpha_{r, a}^*} - \P_0 \phi_{\theta_0, \alpha_{r, a}^*} = \underbrace{(\P_n - \P_0)(\phi_{\hat \theta, \alpha_{r, a}^*} - \phi_{\theta_0, \alpha_{r, a}^*})}_{\text{I}} + \underbrace{\P_0 (\phi_{\hat \theta, \alpha_{r, a}^*} - \phi_{\theta_0, \alpha_{r, a}^*})}_{\text{II}} + \underbrace{(\P_n - \P_0)\phi_{\theta_0, \alpha_{r, a}^*}}_{\text{III}}
\end{align*}
Now for term (I), we may use Lemma 19.24 of \cite{van2000asymptotic} to prove that term I is $o_P(n^{-1/2})$ under the condition that $\phi_{\theta, \alpha_{r, a}^*}$ lies in a Donsker class. For term III,  it is simply $o_P(1)$ by weak law of large numbers. For term II, it is also $o_P(1)$ as $\hat \theta \rightarrow_p \theta$. Thus, this implies that 
$$
\P_n \phi_{\hat \theta, \alpha_{r, a}^*} = o_P(1) + \P_0 \phi_{\theta_0, \alpha_{r, a}^*}
$$
Thus, put everything together, we have that 
\begin{align*}
  \sqrt{n} \P_n \psi_{\hat \theta, \alpha^*} + \sum_{r, a \neq \1_d} \P_0 \phi_{\theta_0, \alpha_{r, a}^*} \sqrt{n} \P_n \xi_{r, a} + o_P(1) = 0
\end{align*}
Next, we have 
\begin{align*}
  \sqrt{n}\P_n \psi_{\hat \theta, \alpha^*} & = \sqrt{n} (\P_n \psi_{\hat \theta, \alpha^*} - \P_0 \psi_{\theta_0, \alpha^*}) \\
  & = \sqrt{n}(\P_n - \P_0)(\psi_{\hat \theta, \alpha^*} - \psi_{\theta_0, \alpha^*}) + \sqrt{n}\P_0 (\psi_{\hat \theta, \alpha^*} - \psi_{\theta_0, \alpha^*}) + \sqrt{n} (\P_n - \P_0)\psi_{\theta_0, \alpha^*} \\
  & = o_P(1) + \sqrt{n} \nabla_{\theta} \P_0 \psi_{\theta_0, \alpha^*} (\hat \theta - \theta_0) + \sqrt{n}(\P_n - \P_0)\psi_{\theta_0, \alpha^*}
\end{align*}

Next, put everything together, we have 
\begin{align*}
  \sqrt{n} \nabla_\theta \P_0 \psi_{\theta_0, \alpha^*} (\hat \theta - \theta_0) = -\sqrt{n}(\P_n - \P_0)\psi_{\theta_0, \alpha^*} - \sum_{r, a \neq \1_d} \P_0 \phi_{\theta_0, \alpha_{r, a}^*} \sqrt{n} \P_n \xi_{r, a} + o_P(1)
\end{align*}
which implies that 
\begin{align*}
  \sqrt{n} (\hat \theta - \theta_0) = - (\nabla_\theta \P_0 \psi_{\theta_0, \alpha^*})^{-1} \left[ \sqrt{n}(\P_n - \P_0)\psi_{\theta_0, \alpha^*} + \sum_{r, a \neq \1_d} \P_0 \phi_{\theta_0, \alpha_{r, a}^*} \sqrt{n} \P_n \xi_{r, a} \right] + o_P(1)
\end{align*}
Thus, we have the desired asymptotic normality for $\hat \theta$. 
\end{proof}

\section{Derivations for Simulation Studies} \label{sec::app::simulation}


We first derive $\E[Y_3]$ with regression adjustment for single variables case. We have that 
\begin{align*}
p(y_3 | A = 1, R \geq 00) = \frac{P(y_3, A = 1, R \geq 00)}{P(A = 1, R \geq 00)} = \frac{3}{4} \phi_{1, 1}(y_3) + \frac{1}{4} \phi_{0, 1}(y_3)
\end{align*}
Thus, we have that $E[Y_3 | A = 1, R \geq 00] = 3/4$. Under ACCMV assumption, we can then identify $P(y_3 | A = 0, R = 00)$ as follows:
$$
P(y_3 | A = 0, R = 00) = p(y_3 | A = 1, R \geq 00) = \frac{3}{4} \phi_{1, 1}(y_3) + \frac{1}{4} \phi_{0, 1}(y_3)
$$
Next, we have that 
\begin{align*}
  P(y_3 | A = 1, R \geq 01, y_2) = \frac{P(y_3, A = 1, R \geq 01, y_2)}{P(A = 1, R \geq 01, y_2)}
\end{align*}
Under ACCMV assumption, we have that 
$$
P(y_3 | A = 1, R = 01, y_2) = P(y_3 | A = 1, R \geq 01, y_2) = \frac{1}{2}\left(\frac{\phi_{\mu_2, \Sigma_2}(y_3, y_2)}{\phi_{-1, 1}(y_2)} + \frac{\phi_{\mu_4, \Sigma_2}(y_3, y_2)}{\phi_{-1, 1}(y_2)} \right)
$$
with $\mu_4 = (0, -1)^T$.
Then we can compute that 
\begin{align*}
  E[Y_3 | A = 1, R \geq 01, Y_2] =  \frac{Y_2}{2} + 1
\end{align*} 
Similarly, we have that 
\begin{align*}
  E(Y_3 | A = 1, R \geq 10, Y_1) = \frac{Y_1}{2} + 1
\end{align*}
Finally, we also have that 
\begin{align*}
  E(Y_3 | A = 1, R = 11, Y_1, Y_2) = \frac{1}{3}(Y_1 + Y_2) + \frac{2}{3}
\end{align*}
Thus, we could compute the parameter of interest $\E[Y_3]$ as 
\begin{align*}
  E[Y_3] & = \E[Y_3 I(A = 1)] + \sum_{r} \E[ m_{r,0}(X_r) I(R = r, A = 0)]
\end{align*}
where $\E[Y_3 I(A = 1)] = \E[Y_3 I(A = 1, R \geq 00)] = 3/8$ and 
\begin{align*}
  & \E[m_{00, 0}(X_{00})I(R = 00, A = 0)] = \frac{3}{32} \\
  & \E[m_{10, 0}(X_{10})I(R = 10, A = 0)] = \frac{1}{2}\E[Y_1 I(A = 0, R = 01)] + \frac{1}{8} = \frac{3}{16} \\
  & \E[m_{01, 0}(X_{01})I(R = 01, A = 0)] = \frac{1}{2}\E[Y_2 I(A = 0, R = 10)] + \frac{1}{8} = \frac{3}{16} \\
  & \E[m_{11, 0}(X_{11})I(R = 11, A = 0)] = \frac{1}{3}\E[(Y_1 + Y_2) I(A = 0, R = 11)] + \frac{1}{12} = \frac{1}{12}
\end{align*}

For IPW estimation of $\E[Y_3]$, we first have that 
\begin{align*}
  O_{00} = \frac{P(A = 0, R = 00)}{P(A = 1, R \geq 00)} = \frac{1}{4}
\end{align*}
Further, when $r = 10$, we have 
\begin{align*}
  O_{10}(y_1) = \frac{P(A = 0, R = 10 | y_1)}{P(A = 1, R \geq 10 | y_1)} = \frac{P(y_1, A = 0, R = 10)}{P(y_1, A = 1, R \geq 10)} = \frac{1}{2}\exp(2y_1)
\end{align*}
Similarly, we have that 
\begin{align*}
  O_{01}(y_2) = \frac{P(A = 0, R = 01 | y_2)}{P(A = 1, R \geq 01 | y_2)} = \frac{P(y_2, A = 0, R = 01)}{P(y_2, A = 1, R \geq 01)} = \frac{1}{2} \exp(2 y_2)
\end{align*}
Finally, for $r = 11$, we have that 
\begin{align*}
  O_{11}(y_1, y_2) &= \frac{P(A = 0, R = 11 | y_1, y_2)}{P(A = 1, R = 11 | y_1, y_2)} = \frac{P(y_1, y_2, A = 0, R = 11)}{P(y_1, y_2, A = 1, R = 11)} \\
  &= \frac{\phi_{\mu_1, \Sigma_1}(y_1, y_2)}{\phi_{(-1, -1), \Sigma_1}(y_1, y_2)} = \exp\left( 
    \frac{8}{3}y_1 - \frac{4}{3}y_2 - \frac{4}{3}
  \right)
\end{align*}

Now we move to the case of multiple primary variables and derive $\E[Y_3 Y_4]$. We have that 
\begin{align*}
  P(y_3, y_4 | A = 00, R = 00) = P(y_3, y_4 | A = 11, R \geq 00) = \phi_{\1_2, \Sigma_2}(y_3, y_4)
\end{align*}
Then we have 
\begin{align*}
  \E[Y_3 Y_4 | A = 11, R \geq 00] = 3/2
\end{align*}
Next, we have 
\begin{align*}
  P(y_3, y_4 | A = 00, R = 01, y_2) = P(y_3, y_4 | A = 11, R \geq 01, y_2) = \frac{P(y_2, y_3, y_4, A = 11, R \geq 01)}{P(y_2, A = 11, R \geq 01)}
\end{align*}
and 
\begin{align*}
  Y_3, Y_4 | A = 11, R \geq 01, Y_2 \sim N\left( \left(\begin{array}{c} \frac{1}{2}Y_2 + \frac{1}{2} \\ \frac{1}{2}Y_2 + \frac{1}{2} \end{array}\right), 
  \left(\begin{array}{cc} 3/4 & 1/4 \\ 1/4 & 3/4 \end{array}\right) \right)
\end{align*}
Then we have that 
\begin{align*}
  \E[Y_3 Y_4 | A = 11, R \geq 01, Y_2] = \frac{1}{4} + \left(\frac{1}{2}Y_2 + \frac{1}{2}\right)^2
\end{align*}
Next, we have 
\begin{align*}
  P(y_3, y_4 | A = 00, R = 10, y_1) = P(y_3, y_4 | A = 11, R \geq 10, y_1) = \frac{P(y_1, y_3, y_4, A = 11, R \geq 10)}{P(y_1, A = 11, R \geq 10)}
\end{align*}
Thus, we now have that 
\begin{align*}
  \E[Y_3 Y_4 | A = 11, R \geq 10, Y_1] = \frac{1}{4} + \left(\frac{1}{2} Y_1 + \frac{1}{2}\right)^2
\end{align*}
Next, we have that 
\begin{align*}
  P(y_3, y_4 | A = 00, R = 11, y_1, y_2) = P(y_3, y_4 | A = 11, R = 11, y_1, y_2) = \frac{\phi_{\1_4, \Sigma_4}(y_1, y_2, y_3, y_4)}{\phi_{\1_2, \Sigma_2}(y_1, y_2)}
\end{align*}
Thus, we have that 
$$
Y_3, Y_4 | A = 11, R = 11, Y_1, Y_2 \sim N\left( \left(\begin{array}{c} \frac{1}{3}(Y_1 + Y_2 + 1) \\ \frac{1}{3}(Y_1 + Y_2 + 1) \end{array}\right), 
  \left(\begin{array}{cc} 2/3 & 1/6 \\ 1/6 & 2/3 \end{array}\right) \right)
$$
and 
\begin{align*}
  \E[Y_3 Y_4 | A = 11, R = 11, Y_1, Y_2] = \frac{1}{6} + \frac{1}{9}(Y_1 + Y_2 + 1)^2
\end{align*}
Next consider the case $A = 01$, we have 
\begin{align*}
  P(y_3 | y_4, A = 01, R = 00) = p(y_3 | y_4, A = 11, R \geq 00) = \frac{p(y_3, y_4, A = 11, R \geq 00)}{p(y_4, A = 11, R \geq 00)}
\end{align*}
Thus, we have that 
\begin{align*}
  Y_3 | Y_4, A = 11, R \geq 00 \sim N\left(\frac{1}{2} Y_4 + \frac{1}{2}, \frac{3}{4}\right)
\end{align*}
and then 
\begin{align*}
  \E[Y_3 Y_4 | Y_4, R \geq 00, A = 11] = Y_4 \E[Y_3 | Y_4, R \geq 00, A = 11] = \frac{1}{2} Y_4 (Y_4 + 1)
\end{align*}
Next, we have 
\begin{align*}
  P(y_3 | y_4, A = 01, R = 01, y_2) = P(y_3 | y_2, y_4, A = 11, R \geq 01) = \frac{P(y_2, y_3, y_4, A = 11, R \geq 01)}{P(y_2, y_4, A = 11, R \geq 01)}
\end{align*}
Then we have 
\begin{align*}
  Y_3 | Y_2, Y_4, A = 11, R \geq 01 \sim N\left( \frac{1}{3}(Y_2 + Y_4 + 1), \frac{2}{3} \right)
\end{align*}
Thus, we have 
\begin{align*}
  \E[Y_3 Y_4 | Y_2, Y_4, A = 11, R \geq 01] = \frac{1}{3}Y_4(Y_2 + Y_4 + 1) 
\end{align*}
Similarly, we have 
\begin{align*}
  P(y_3 | y_4, A = 01, R = 10, y_1) = P(y_3 | y_4, A = 11, R \geq 10, y_1)
\end{align*}
and we can get that $Y_3 | Y_1, Y_4, A = 11, R \geq 10 \sim N\left(
\frac{1}{3}(Y_1 + Y_4 + 1), \frac{2}{3}
\right)$. Thus, we have 
\begin{align*}
  \E[Y_3 Y_4 | Y_1, Y_4, A = 11, R \geq 10] = \frac{1}{3}Y_4(Y_1 + Y_4 + 1)
\end{align*}
Next, we have 
\begin{align*}
  P(y_3 | y_4, A = 01, R = 11, y_1, y_2) = P(y_3 | y_4, A = 11, R = 11, y_1, y_2) = \frac{P(y_1, y_2, y_3, y_4, A = 11, R = 11)}{P(y_1, y_2, y_4, A = 11, R = 11)}
\end{align*}
and we can get that 
\begin{align*}
  Y_3 | Y_4, A = 11, R = 11, Y_1, Y_2 \sim N\left(
    \frac{1}{4}(Y_1 + Y_2 + Y_4 + 1), \frac{5}{8}
  \right)
\end{align*}
Thus, we have 
\begin{align*}
  \E[Y_3 Y_4 | Y_4, A = 11, R = 11, Y_1, Y_2] = \frac{1}{4}(Y_1 + Y_2 + Y_4 + 1)Y_4
\end{align*}

Next consider the case $A = 10$, we have 
\begin{align*}
  P(y_4 | y_3, A = 10, R = 00) = P(y_4 | y_3, A = 11, R \geq 00)
\end{align*}
By symmetry, we have that $Y_4 | Y_3, A = 11, R \geq 00 \sim N(\frac{1}{2}Y_3 + \frac{1}{2}, \frac{3}{4})$. Thus, we have that 
\begin{align*}
  \E[Y_3 Y_4 | Y_3, A = 11, R \geq 00] = \frac{1}{2}Y_3(Y_3 + 1)
\end{align*}
Next we have 
\begin{align*}
  P(y_4 | y_3, A = 10, R = 01, y_2) = P(y_4 | y_2, y_3, A = 11, R \geq 01)
\end{align*}
Again similarly, we have that $Y_4 | Y_2, Y_3, A = 11, R \geq 01 \sim N\left(\frac{1}{3}(Y_2 + Y_3 + 1), \frac{2}{3}\right)$. Thus, we have that 
\begin{align*}
  \E[Y_3 Y_4 | Y_2, Y_3, A = 11, R \geq 01] = \frac{1}{3}Y_3(Y_2 + Y_3 + 1)
\end{align*}
Next, we have 
\begin{align*}
  P(y_4 | y_3, A = 10, R = 10, y_1) = P(y_4 | y_3, A = 11, R \geq 10, y_1)
\end{align*}
and similarly we have $Y_4 | Y_1, Y_3, A = 11, R \geq 10 \sim N\left(
  \frac{1}{3}(Y_1 + Y_3 + 1), \frac{2}{3}
\right)$. Thus, we have that 
\begin{align*}
  \E[Y_3 Y_4 | Y_1, Y_3, A = 11, R \geq 10] = \frac{1}{3}Y_3(Y_1 + Y_3 + 1) 
\end{align*}
Next, we have that 
\begin{align*}
  P(y_4 | y_3, A = 10, R = 11, y_1, y_2) =P(y_4 | y_1, y_2, y_3, A = 11, R = 11)
\end{align*}
and similarly we have $Y_4 | Y_1, Y_2, Y_3, A = 11, R = 11 \sim N\left(
  \frac{1}{4}(Y_1 + Y_2 + Y_3 + 1), \frac{5}{8}
\right)$ and we have that 
\begin{align*}
  \E[Y_3 Y_4 | Y_1, Y_2, Y_3, A = 11, R = 11] = \frac{1}{4} Y_3 (Y_1 + Y_2 + Y_3 + 1)
\end{align*}
Thus, we could now compute the parameter of interest $\E[Y_3 Y_4]$ as 
\begin{align*}
  \E[Y_3 Y_4] = \E[Y_3 Y_4 I(A = 11)] + \sum_{r, a \neq 11} \E[Y_3 Y_4 I(A = a, R = r)]
\end{align*}
where 
\begin{align*}
  \E[Y_3 Y_4 I(A = 11)] = \E[
    \E[Y_3 Y_4 | A = 11] I(A = 11)
  ] = \frac{3}{2} * P(A = 11) = \frac{3}{8}
\end{align*}
Next, when $a = 00$, we have 
\begin{align*}
 & \E[Y_3 Y_4  I(A = 00, R = 00)] = \E[
  m_{00, 00}(X_{00}, L_{00}) I(A = 00, R = 00)
 ] = \frac{3}{2} \times \frac{1}{16} = \frac{3}{32} \\
 & \E[Y_3 Y_4 I(A = 00, R = 01)] = \E[
   m_{01, 00}(X_{01}, L_{00})I(A = 00, R = 01)
 ] = \frac{17}{256} \\
 & \E[Y_3 Y_4 I(A = 00, R = 10)] = \E[m_{10, 00}(X_{10}, L_{00})I(A = 00, R = 10)] = \frac{17}{256}\\
 & \E[Y_3 Y_4 I(A = 00, R = 11)] = \E[m_{11, 00}(X_{11}, L_{00})I(A = 00, R = 11)] = \frac{3}{32}
\end{align*}
Next, when $a = 01$, we have 
\begin{align*}
 & \E[Y_3 Y_4 I(A = 01, R = 00)] = \E[m_{00, 01}(X_{00}, L_{01}) I(A = 01, R = 00)] = \frac{7}{128} \\
 & \E[Y_3 Y_4 I(A = 01, R = 01)] = \E[m_{01, 01}(X_{01}, L_{01}) I(A = 01, R = 01)] = \frac{3}{32} \\
 & \E[Y_3 Y_4 I(A = 01, R = 10)] = \E[m_{10, 01}(X_{10}, L_{01}) I(A = 01, R = 10)] = \frac{3}{32} \\
 & \E[Y_3 Y_4 I(A = 01, R = 11)] = \E[m_{11, 01}(X_{11}, L_{01}) I(A = 01, R = 11)] = \frac{3}{32}
\end{align*}
Finally, the results for $a = 10$ are identical to $a = 01$. Thus, collecting all the terms, we can get that $\E[Y_3 Y_4] = \frac{175}{128}$. 


Now we prove that the simulation setup for the marginal parametric model satisfies the ACCMV assumption.  For $a \neq 11$, we have 
\begin{align*}
  P(R = 1, A = a | X, L) = \frac{\exp(0.5 X)}{5 + 3 \exp(0.5 X)}
\end{align*}
and we have 
$$
P(R = 1, A = 11 | X, L) = \frac{1}{5 + 3 \exp(0.5 X)}
$$
Thus for $a \neq 11$, 
\begin{align*}
  \frac{P(R = 1, A = a | X, L)}{P(R = 1, A = 11 | X, L)} = \exp(0.5 X)
\end{align*} 
which does not depend on $L$. Next, for $R = 0$, we have 
\begin{align*}
  P(R = 0, A = a, x, \ell) = P(R = 0, A = a | x, \ell) f_{X, L}(x, \ell) = \frac{1}{5 + 3 \exp(0.5 x)} f_{X, L}(x, \ell)
\end{align*}
and $f_{X, L}(x, \ell)$ is the density function for $X, L$. Thus, we have 
\begin{align*}
  & P(R = 0, A = a, \ell) = \int \frac{1}{5 + 3 \exp(0.5 x)} f_{X, L}(x, \ell) dx = \int \frac{1}{5 + 3 \exp(0.5 x)} f_{X | L}(x | \ell) dx f_L(\ell) \\
  & \Leftrightarrow P(R = 0, A = a | \ell) = \int \frac{1}{5 + 3 \exp(0.5 x)} f_{X | L}(x | \ell) dx
\end{align*}
and this holds for all $a$. Similarly, we have 
\begin{align*}
  P(R = 1, A = 11 | \ell) = \int \frac{1}{5 + 3 \exp(0.5 x)} f_{X | L}(x | \ell) dx
\end{align*}
Thus, for any $a \neq 11$, 
\begin{align*}
\frac{P(R = 0, A = a | \ell)}{P(R \geq 0, A = 11 | \ell)} = \frac{1}{2}
\end{align*}



\vskip 0.2in
\bibliography{gang}

\end{document}